\documentclass[a4paper,12pt]{article}
\usepackage{graphicx}

\usepackage{amssymb}
\usepackage{amsmath}
\def\UP{\ensuremath{U_{\rm p}}}
\def\Urms{\ensuremath{U_{\rm rms}}}
\def\RE{\ensuremath{R}}
\def\RT{\ensuremath{\RE_{\rm t}}}
\def\RG{\ensuremath{\RE_{\rm g}}}
\def\RU{\ensuremath{\RE_{\rm u}}}

\newcommand{\p}{\partial}
\newcommand{\BE}{\begin{equation}}
\newcommand{\EE}{\end{equation}}
\newcommand{\BC}{\begin{center}}
\newcommand{\EC}{\end{center}}
\newcommand{\BF}{\begin{figure}}
\newcommand{\EF}{\end{figure}}

\let\TW=\textwidth

\title{From temporal to spatiotemporal dynamics\\
 in transitional plane Couette flow}

\author{Jimmy Philip and Paul Manneville\\
Laboratoire d'Hydrodynamique\\
Ecole Polytechnique, 91128 Palaiseau, France}

\date{accepted for publication in Phys. Rev. E (2011/01/27)}

\begin{document}

\sloppy
\maketitle

\begin{abstract}
Laboratory experiments point out the existence of patterns made of alternately laminar and turbulent oblique bands in plane Couette flow in its way to/from turbulence as the Reynolds number $R$ is varied.
Many previous theoretical and numerical works on the problem have considered small aspect-ratio systems subjected to periodic boundary conditions, while experiments correspond to the opposite limit of large aspect-ratio.
Here, by means of fully resolved direct numerical simulations of the Navier--Stokes equations at decreasing $R$, we scrutinize the transition from temporal to spatiotemporal behavior in systems of intermediate sizes.
We show that there exists a streamwise crossover size of order $L_x\sim$70--80$\,h$ (where $2h$ is the gap between the plates driving the flow) beyond which the transition to/from turbulence in PCF is undoubtedly a spatiotemporal process, with typical scenario `[turbulent flow]$\>\to\>$[riddled regime]$\>\to\>$[oblique pattern]$\>\to\>$ [laminar flow]', whereas below that size it is more a temporal process describable in terms of finite-dimensional dynamical systems with scenario `[chaotic flow]$\>\to\>$[laminar flow]' ({\it via\/} chaotic transients).
In the crossover region, the `[oblique pattern]' stage is skipped, which leads us to suggest that an appropriate rendering of the patterns observed in experiments needs a faithful account of streamwise correlations at scales at least of the order of that crossover size.

\end{abstract}

\section{Introduction\label{1}}

During the last few years there has been a resurgence of interest regarding the formation of laminar--turbulent patterns in wall-bounded shear flows having very large aspect ratios (when the lateral dimensions, along $x$ and $z$, are more than two orders of magnitude larger than the relevant wall-normal dimension, along $y$). 
The phenomenon was discovered in circular Couette flow (CCF) by Coles and Van Atta \cite{Coles65,CoVa66} who called it `spiral turbulence'.
The pattern can be observed when the two cylinders of CCF rotate in opposite directions in a specific velocity range, which is only a part of the complete bifurcation diagram \cite{An_etal86}.
Only one or two stripes of the laminar--turbulent pattern were observed in those experiments owing to their low aspect ratio (the ratio of the perimeter to the gap between the cylinders).
It is only later that Prigent {\it et al.} \cite{prigent03} made detailed measurements of the relevant features of this phenomenon in their very-large aspect ratio setup, where about
10--15 stripes could be obtained.

A similar event of laminar--turbulent pattern formation was also observed by the same team \cite{prigent03} in large-aspect-ratio plane Couette flow (PCF), a schematic of which is shown in the top panel of figure \ref{fig1}, the bottom one displaying a snapshot of the pattern obtained
experimentally.
When the control parameters of CCF and PCF are made equivalent by using appropriate scales, both CCF and PCF display patterns in almost the same parameter range \cite{Manneville04}, with the obvious difference of streamwise periodicity present in CCF and not in PCF.
In the following we leave aside the case of CCF whose bifurcation diagram is slightly more complicated owing to the interplay of centrifugal instability mechanisms and concentrate our attention on PCF in the range of parameters relevant to the transition to turbulence.
\BF
\BC
\includegraphics[width=0.65\TW,clip]{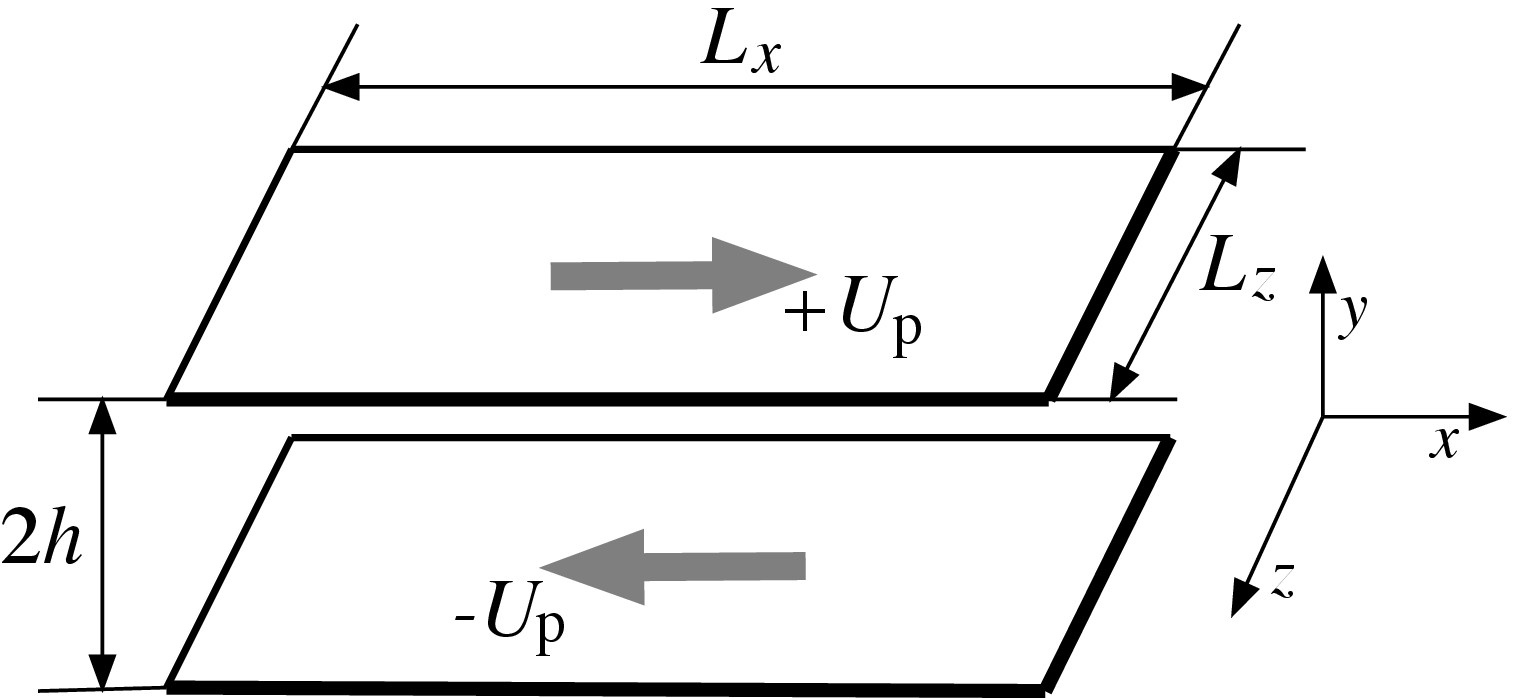}\\[4ex]
\includegraphics[width=0.8\TW,clip]{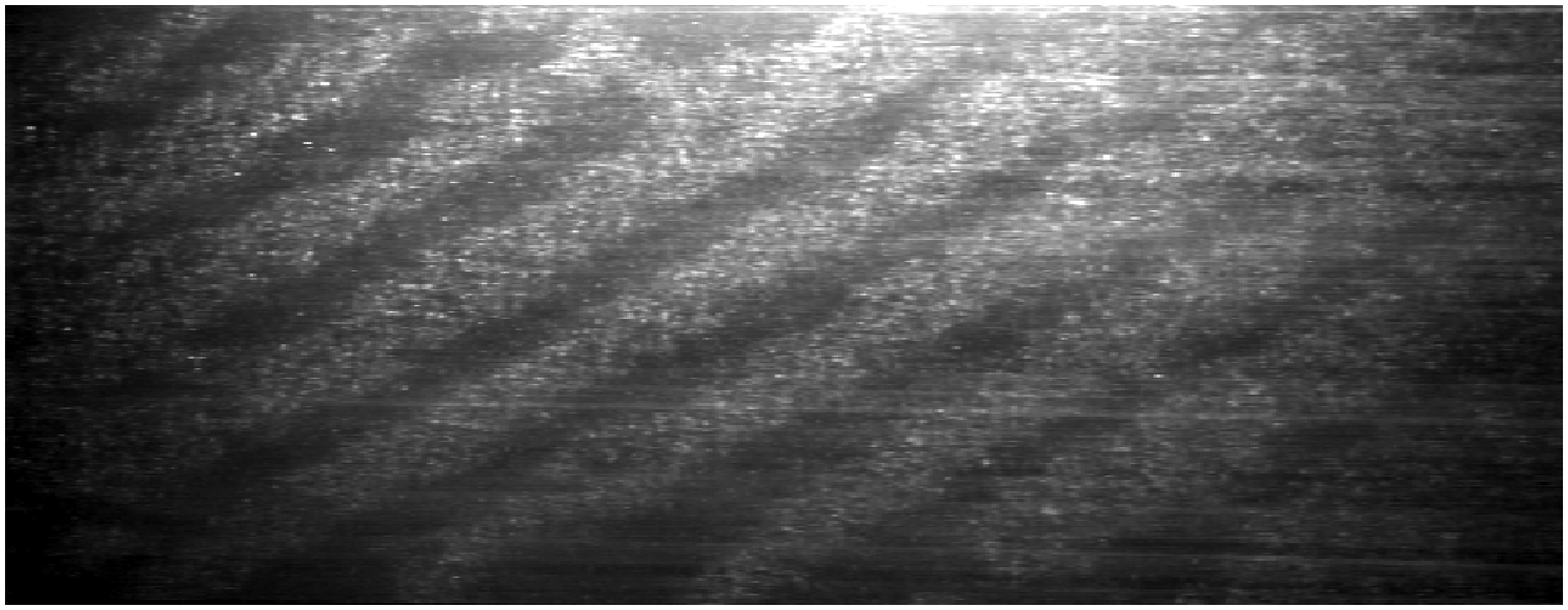}
\EC
\caption{Top: Schematic of PCF with top and bottom plate moving at $\pm\UP$ (at $y=\pm h$) separated by a distance $2h$.
Bottom: Experimental visualization of laminar(dark)--turbulent(light) pattern
in PCF at $\RE=358$, $L_x=770\, h$, $L_z=430\, h$ (courtesy A. Prigent).
For comparison, most of S.~Bottin's experiments were performed with $L_x=284\,h$, $L_z=72.6\,h$ \cite{Betal98,Bo98}.
The size of the MFU is $\lambda_x=2\pi/\alpha$, $\lambda_z=2\pi/\gamma$, where $\alpha$ and $\gamma$ are the fundamental wavenumbers in the streamwise and spanwise directions, respectively.
Here we take the values corresponding to the self-sustained exact solution found by Waleffe \cite{Waleffe03},  $\alpha=0.49$ and $\gamma=1.5$ and thus $\lambda_x=12.8$ and $\lambda_z=4.2$.
\label{fig1}}
\EF

Traditionally, the control parameter is the Reynolds number, and for PCF it is defined as $\RE:=\UP h/\nu$, here based on the half channel height $h$, the plate velocity \UP, and the kinematic viscosity $\nu$;
in the following $h$ and the advection time $\tau_{\rm a}=h/\UP$ are used as length and time units, respectively.
(The viscous relaxation time is $\tau_{\rm v}=h^2/\nu$, so that $\RE=\tau_{\rm v}/\tau_{\rm a}$.)
The transition process can be studied by increasing or decreasing $R$.
The essential point is that the laminar flow is stable against infinitesimal perturbations for all $R$, so that a direct transition to turbulence is observed when finite amplitude perturbations are introduced, according to a {\it globally subcritical\/} scenario (for an introductory review
see~\cite[Chap.7]{Ma10}).
Studies at increasing $R$ are strongly sensitive to the amplitude and shape of the triggering perturbation.
Quantitatively different results about the transition can accordingly be obtained under different protocols, e.g. \cite{TiAl92,Bo98}, while the picture of course remains qualitatively unchanged.
In contrast, the experiments of Prigent {\it et al.} (see \cite{Pr01} for details), were systematically performed by varying $R$ in small steps while waiting for statistical equilibrium at each $R$, which helps us to clearly identify several stages.
First, beyond $\RT\approx410$, turbulence is essentially uniform or `featureless', borrowing the term introduced by Andereck {\it et al.} for CCF \cite{An_etal86}.
Next, oblique laminar--turbulent bands appear upon decreasing \RE\ slowly below \RT.
The amplitude of the  laminar--turbulent modulation grows continuously as \RE\ is further decreased, which has been interpreted as a {\it supercritical\/} bifurcation in the presence of a noise reminiscent of featureless turbulence \cite{prigent03}.
The bands next become fragmented and turn into irregular oblique turbulent patches which seem sustained for $\RE\ge\RG\approx325$ but decay in a finite time for $\RE<\RG$.
The value of \RG\ mentioned above has also been obtained in experiments where turbulent spots were triggered \cite{Betal98,Bo98}.
Below \RG\ the lifetimes of turbulent spots are distributed according to decreasing exponentials whose characteristic time is seen to diverge as \RE\ approaches \RG\ from below \cite{Betal98}.
Finally, for $\RE<\RU\approx280$ large perturbations relax without measurable waiting time and in a mostly monotonic way.

Direct numerical simulation (DNS) of these large aspect ratio systems was delayed because of the huge computational requirements.
Barkley and Tuckerman \cite{BarTuc05} were the first to obtain the band patterns in fully resolved simulations of the Navier--Stokes equations, however in carefully chosen narrow and tilted computational domains.
Their choice of domain however precluded the occurrence of patterns with defects or orientation changes inside the flow.
This restriction was overcome in the DNS by Duguet {\it et al.} \cite{DugSchDan10} who recovered the experimental findings of Prigent {\it et al.} in a fully resolved very large aspect ratio system.
Similarly, the spiral regime of CCF was numerically obtained by Meseguer {\it et al.} \cite{MeMeAvMa09} and Dong \cite{Dong09}.
An oblique band pattern was also found numerically by Tsukahara {\it et al.} \cite{TsSeKaTo05} in the case of plane Poiseuille flow (PPF).
The above three systems, CCF, PCF and PPF are prototypical for the study of laminar--turbulent patterns; all of them are confined by two walls in direction $y$, whereas the other two dimensions extend to infinity, in the limit.

Two other flows of theoretical and technological importance showing behavior similar to the above-mentioned systems are pipe flow and the flat-plate boundary-layer flow.
Pipe flow is in some sense simpler, owing to only one spatially-extended dimension, the length of the pipe.
Inside a specific Reynolds number range, localized patches of turbulence (`puffs') are observed separated by laminar regions.
At present it is not clear how they are related to the oblique bands in other systems \cite{MoxBar10}.
On the other hand, the boundary layer flow case gets more complicated due to the fact that the relevant wall normal dimension, the boundary-layer thickness, increases with the flow direction and, with it, the relevant value of \RE.
Turbulent spots that appear amid laminar flow during the transition are however strikingly similar to those growing into oblique bands observed in PCF.

In contrast with instabilities in {\it closed flow\/} systems such as Rayleigh--B\'enard convection for which the aspect ratio is a genuine control parameter that can be varied from small to large by changing the position of walls in all directions \cite[Chap.3]{Ma10}, {\it open flows\/} develop most often in domains physically less constrained by lateral walls, while streamwise boundary conditions make it reasonable to accept the assumption of translational invariance in that direction, at least locally.
In this respect, all of the investigations mentioned above, both experimental and numerical, are related to large aspect ratio systems and represent the typical situation, even if the experimental setup is difficult to construct \cite{TiAl92,Bo98,prigent03} or the DNS computationally demanding \cite{DugSchDan10,TsSeKaTo05}.
Despite its relevance, this is {\it not\/} the situation that has been considered in many recent works on the transitional problem.
Indeed, most of the theoretical and numerical work has been performed in configurations confined by periodic boundary conditions at small distances.
Seminal work in the 1990's bore on high resolution simulations devoted to the identification of the so-called  {\it minimal flow unit\/} (MFU) below which turbulence can not be sustained, as done by Jimenez and Moin \cite{JimMoi91}, later used for the elucidation of the {\it self-sustaining process\/} (SSP), the mechanism of sustenance of turbulence by which streamwise vortices induce streaks that break down to regenerate the vortices, following Hamilton {\it et al.\/} \cite{HamKimWal95}, next to the discovery of exact coherent states of Navier--Stokes equations,
following Nagata \cite{Nagata90} and Waleffe \cite{Waleffe03}, and to the study of the boundary of the attraction basin of the laminar flow \cite{EcFaScSc08}.

In fact setting boundary conditions at small distances reduce the infinite-dimensional dynamical problem posed by the Navier--Stokes equations to a finite-dimensional problem as long as \RE\ remains moderate, which is the case in the transitional regime (but would be insufficient in the high-\RE\ limit where fine vortical structures develop already at the level of the MFU).
Accordingly all the works in small domains to some extent have come under the purview of temporal dynamics and chaos theory.
The investigation of such small systems has indeed been extremely valuable \cite{EcFaScSc08,GiHaCv08}, with important achievements such as the recent findings of `edge states' and `localized solutions' in PCF \cite{DuScHe09,ScMaEc10}, as an encouragement to make the connection to special solutions and associated bifurcation structures obtained in pattern-forming model equations \cite{BuKc07,ScGiBu10}. 

However, any accurate representation of the dynamics at the level of the MFU, though remaining instructive, is not informative of the experimental situation since the smallest setups that have been used should rather be analyzed as two-dimensional arrays made of tens or
hundreds of MFUs (see caption of Fig.~\ref{fig1}), which allows for global spatiotemporal dynamics, while placing periodic boundary conditions at the scale of the MFU grossly overestimates the coherence of the flow.
The size of the system ($\gg$ MFU) and the coexistence of two possible local states, either laminar or chaotic, each corresponding to a possible temporal regime at the MFU scale, make it possible for whole regions, either laminar or turbulent, to coexists in physical space. The possibility of such modulations roots the recourse to concepts from the theory of spatiotemporal chaos \cite{CrHo93}.  Within this general framework,  the pattern$\>\to\>$featureless transition in PCF, TCF and other similar flows, appears to be a symmetry-restoring bifurcation observed upon increasing \RE\ for which an order parameter can be defined \cite{prigent03,Betal08}, 
with the understanding that the base state is the translationally invariant, strongly noisy, featureless regime beyond \RT.

A previous attempt to reach the spatiotemporal level directly {\it via\/} modeling by one of us~\cite{LaMa07}, though promising~\cite{Ma09}, however failed to reproduce the bands, due to
insufficient wall-normal resolution.
Another avenue to spatiotemporal dynamics is through DNS of the Navier--Stokes equations but, as already mentioned, long duration simulations of wide enough domains is still too demanding.
Reducing the wall-normal resolution has been shown a viable option at a qualitative level \cite{MaRo10} but it was not clear that the quantitative shift observed on the transitional range $[\RG,\RT]$ was without hidden consequence on the pattern formation problem.

The aim of the present work is at studying, by means of fully resolved
DNS of Navier--Stokes equations, what happens when the system size $L$
(to be defined more precisely later) increases and the temporal dynamics
gives way to a spatiotemporal one. The next section anticipates
the outcome of the study before giving details on the numerics.
Section \ref{3} contains the main results extracted from the
numerical simulations as well as the bifurcation diagram. The final
section summarizes the study and draws conclusions.

\section{General framework for the study of transitional PCF\label{2}}

\subsection {Expected bifurcation diagram for the Turbulent--Laminar
transition\label{2_1}}

\BF
\BC
\centerline{\includegraphics[width=1.0\textwidth,clip]{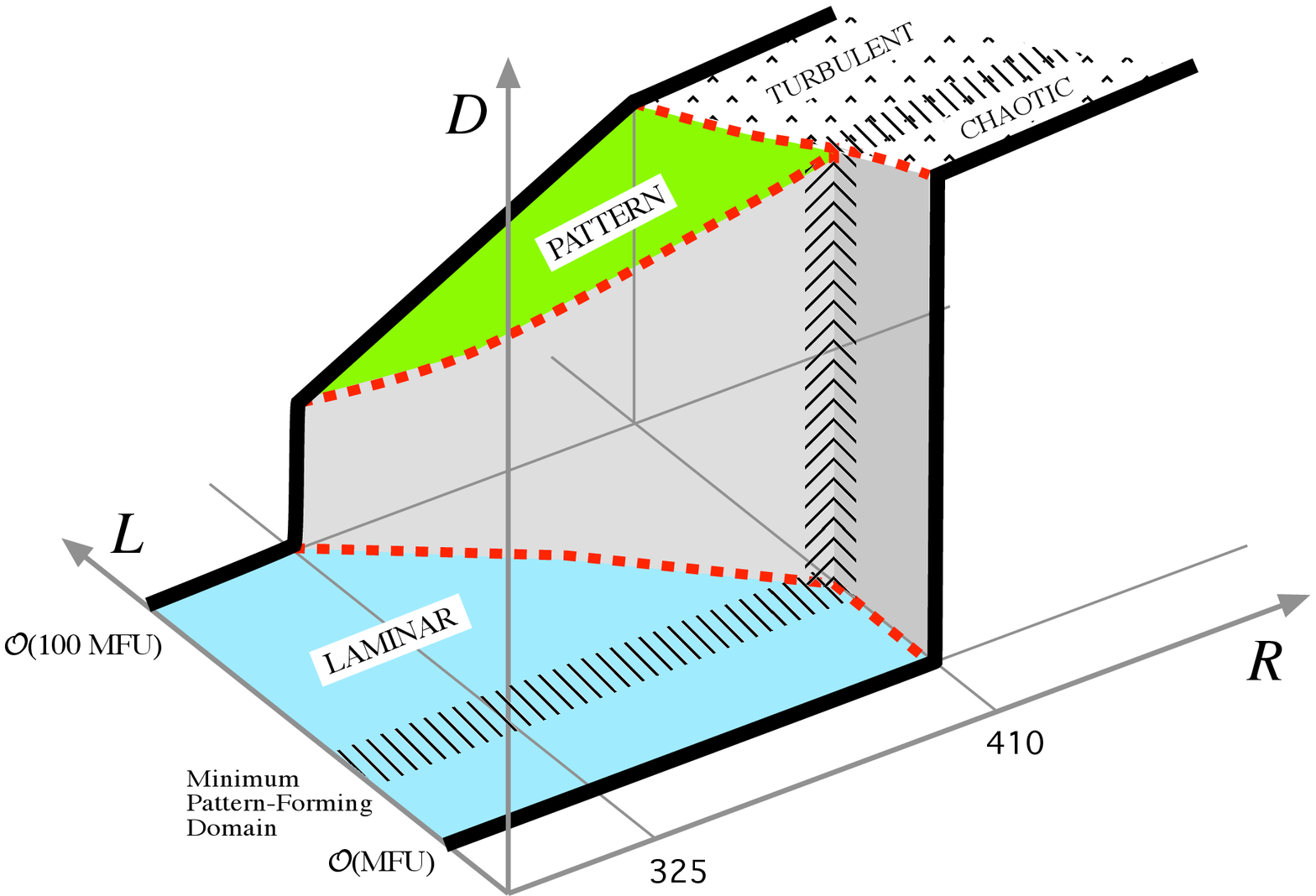}}
\caption{(Color online) Schematic of the expected bifurcation diagram for PCF including
the locations of domains with the size of the order of MFU and
that of pattern-forming systems.\label{fig2}}
\EC
\EF

Figure \ref{fig2} gives a schematic three-dimensional view of the bifurcation diagram for transitional PCF in the two well-studied cases of MFU-like systems and large aspect-ratio systems, and propose an interpolation between these two cases as an educated guess.
We use \RE, $D$,  and $L$ as a coordinate system.
Besides \RE, the Reynolds number, $D$ is a measure of the distance to laminar flow, e.g. the time-averaged, domain-averaged rms of the velocity departure from the laminar flow profile, and $L$ is the characteristic lateral domain size.
The diagram shows two thick black lines in two (\RE--$D$) planes, one at $L={\cal O}$(MFU), the other at $L={\cal O}$(100 MFU) where `100' is just meant to be a large number typical of laboratory experiments.
They sketch the variation of $D$ as a function of $R$, as the latter parameter is gradually reduced from a high value where the system is turbulent.

The first curve, at  $L={\cal O}$(MFU) is typical of a {\it subcritical\/} bifurcation with an abrupt jump from turbulence ($D$ finite and large) to laminar flow ($D=0$).
Here it is just a tentative sketch of the ideal situation where temporal chaos would break down in a single stage {\it via\/} attractor crisis \cite{Ott06}.
This guess is consistent with the observations of exponentially distributed lifetimes of chaotic transients and the divergence of the average lifetime as the putative crisis point is approached \cite{EcFaScSc08}.
In fact, the very existence of a crisis at finite \RE\ (here situated around $\RE=410$) has not been proved so far but this does not change our picture drastically.

In contrast, the second line at $L={\cal O}$(100 MFU) corresponding to large aspect ratio systems is supported by laboratory and computer experiments \cite{prigent03,DugSchDan10}.
As \RE\ is lowered below $\RT\approx410$, the distance $D$ begins to decrease owing to the coexistence of laminar and turbulent regions in variable amounts in the whole domain. Whereas the turbulence intensity seems to decrease only slightly with \RE, the main part of the decrease of $D$, a quantity averaged over the whole surface of the system, has to be attributed to the increase of the laminar fraction, the complementary of the turbulent fraction.
The spatial organization of this laminar--turbulent coexistence is expressed by the term `{\sc pattern}' (henceforth shortened as `P') used in the figure.
At $\RG\approx325$, the distance $D$ drops to zero because the regime observed in the long-time limit is laminar, but this does not mean that one cannot observe transiently turbulent patches with smaller turbulent fraction and exponentially distributed lifetimes \cite{Betal98}.

Thick red dashed lines  in Fig.~\ref{fig2} suggest changes in the diagram as $L$ varies, showing up two regions more: `{\sc laminar}' (`L') and `{\sc turbulent}' (`T') or `{\sc chaotic}' (`C').
When $L$ is large, `T' actually means `featureless turbulence' but, when $L$ is small, `temporal chaos' would better reflect the spatial coherence in the flow, hence the `C'.
This coherence forbids the emergence of sub-domains that could be identified as laminar or turbulent in the system, which is no longer the case when $L$ gets larger.
For reasons of topological continuity, there should be a  crossover size (shown by a hatched band) below which PCF displays two states, `L' and `C', and above which three states, `L', `P', and `T', the hatched band extending into the `C/T' region to mark the change from temporal chaos to spatiotemporal chaos.
It will shown below that, such a bifurcation diagram indeed exists, with however some peculiarities in the `P' region.

\subsection {Numerical simulation details\label{2_2}}

Direct Numerical Simulation (DNS) of the Navier--Stokes equations is carried out using Gibson's well tested, freely available DNS software {\sc ChannelFlow} \cite{channelflow}.
It is a pseudo-spectral code using Chebyshev polynomials in the wall-normal direction with  no-slip conditions at $y_{\rm p}=\pm1$ and in-plane Fourier modes adapted to periodic boundary conditions at distances $L_{x,z}$ in the $(x,z)$ directions.
The number of Chebyshev modes is $N_y=33$, which is well suited to resolve all the relevant modes of turbulent flow in the range of \RE\ studied according to Duguet {\it et al.} \cite{DugSchDan10} (see also note \footnote{This corresponds to an average wall-normal spacing $\Delta y^+ = 1.81$ based on $\RE_\tau=32$; the superscript `+' denotes quantities scaled by the viscous length unit $\nu/u_\tau$, where $u_\tau$ is the friction velocity $\sqrt{\tau_w/\rho}\,$, ${\tau}_w$ being the shear stress at the wall and $\rho$ the fluid density; further, $\RE_\tau$ is defined as $u_\tau h/\nu$ and $\RE_\tau=32$ roughly corresponds to $\RE=420$.}).
The time step is such as to keep the CFL number between 0.4 and 0.6.
Some details of the various domains used in simulations are given in table \ref{table1}, where $N_x$ and $N_z$ are the number of collocation points in $x$ and $z$ directions, respectively, and $L_x$ and $L_z$, the corresponding domain lengths.
Since the 3/2 rule is applied in all cases to remove aliasing, this corresponds to solutions in the Fourier space using $N_{x,z}'=\frac23N_{x,z}$ modes, or equivalently to space steps $L_{x,z}/N_{x,z}'=\frac32L_{x,z}/N_{x,z}$.
Domains are chosen with increasing size, measured here by the diagonal length $L:=\sqrt{L_x^2+L_z^2}$, which will also be used in referring to various simulations.
Moreover, in all cases $\theta:=\tan^{-1}(L_z/L_x)$ is kept between 20--30$^\circ$, consistent with previous investigations \cite{prigent03,BarTuc07}.
For certain domain sizes, extra computations with an increased number of grid points are also carried out to ensure the independence on grid resolution, e.g. for $L=16.9$, $N_x$ was raised to 96, and for $L=65.4$, $(N_x,N_z)$ were raised to (282,128) without any significant quantitative changes in the results.

\begin{table}
\caption{Computational domains and the corresponding grid points}
\hspace{0.05cm}
\label{table1}
\BC
\begin{tabular}{c c c c c c c c c c}
\hline
$L_x, L_z$ & $N_x, N_z$ & $L_x/N_x$ & $L_z/N_z$ & $L$ & $\theta^\circ$ \\
\hline

$5\pi,2\pi$     &  64,32    & 0.25 & 0.19 &  16.9 & 21.8  \\
24,9              &  94,32    & 0.26 & 0.28 &  25.6 & 20.6  \\
32,15             & 128,64    & 0.25 & 0.23 &  35.3 & 25.1  \\
60,26             & 192,96    & 0.31 & 0.27 &  65.4 & 23.4  \\
70,30             & 282,128   & 0.25 & 0.23 &  76.2 & 23.2  \\
80,35             & 384,192   & 0.21 & 0.18 &  87.3 & 23.6  \\
90,40             & 384,192   & 0.23 & 0.21 &  98.5  &24.0 \\
100,45            & 384,192   & 0.26 & 0.23 &  109.6 & 24.2 \\
128,64            & 512,256   & 0.25 & 0.25 &  143.1 & 26.6 \\
\hline
\end{tabular}
\EC
\end{table}

\BF
\BC
\includegraphics[width=0.45\TW,clip]{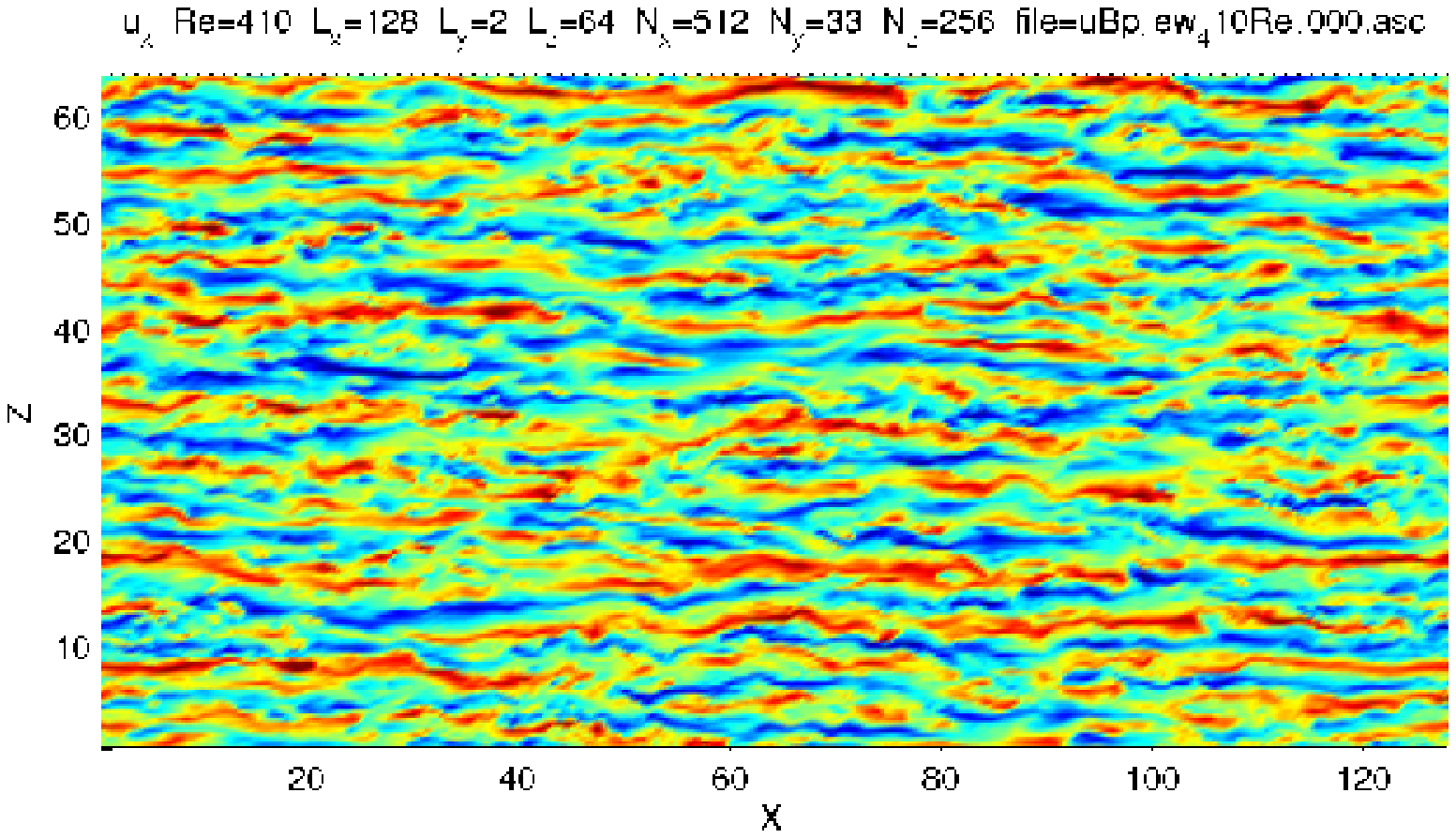}
\hspace{0.05\textwidth}
\includegraphics[width=0.3515625\TW,clip]{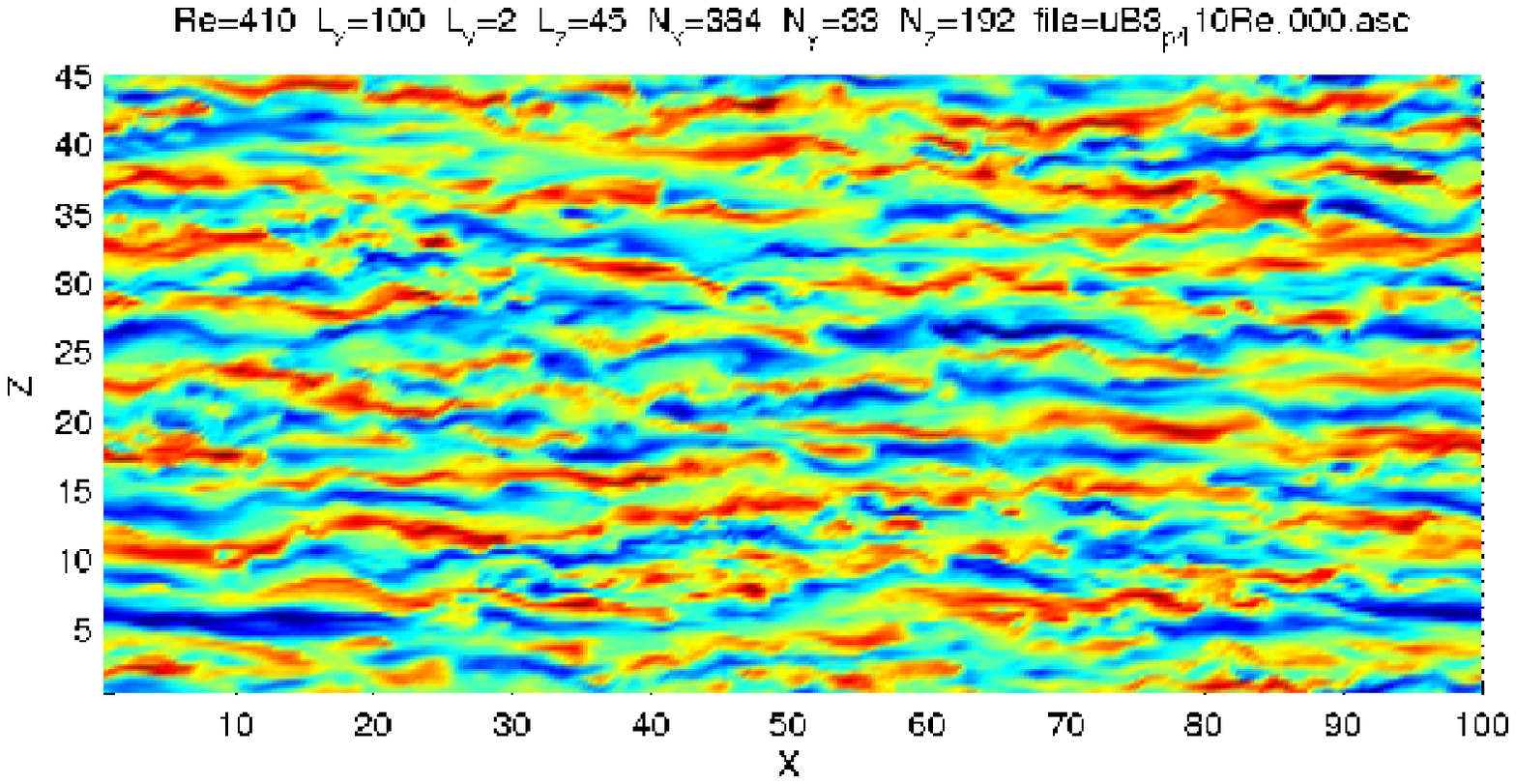}
\vspace{0.5cm}

\includegraphics[width=0.3164\TW,clip]{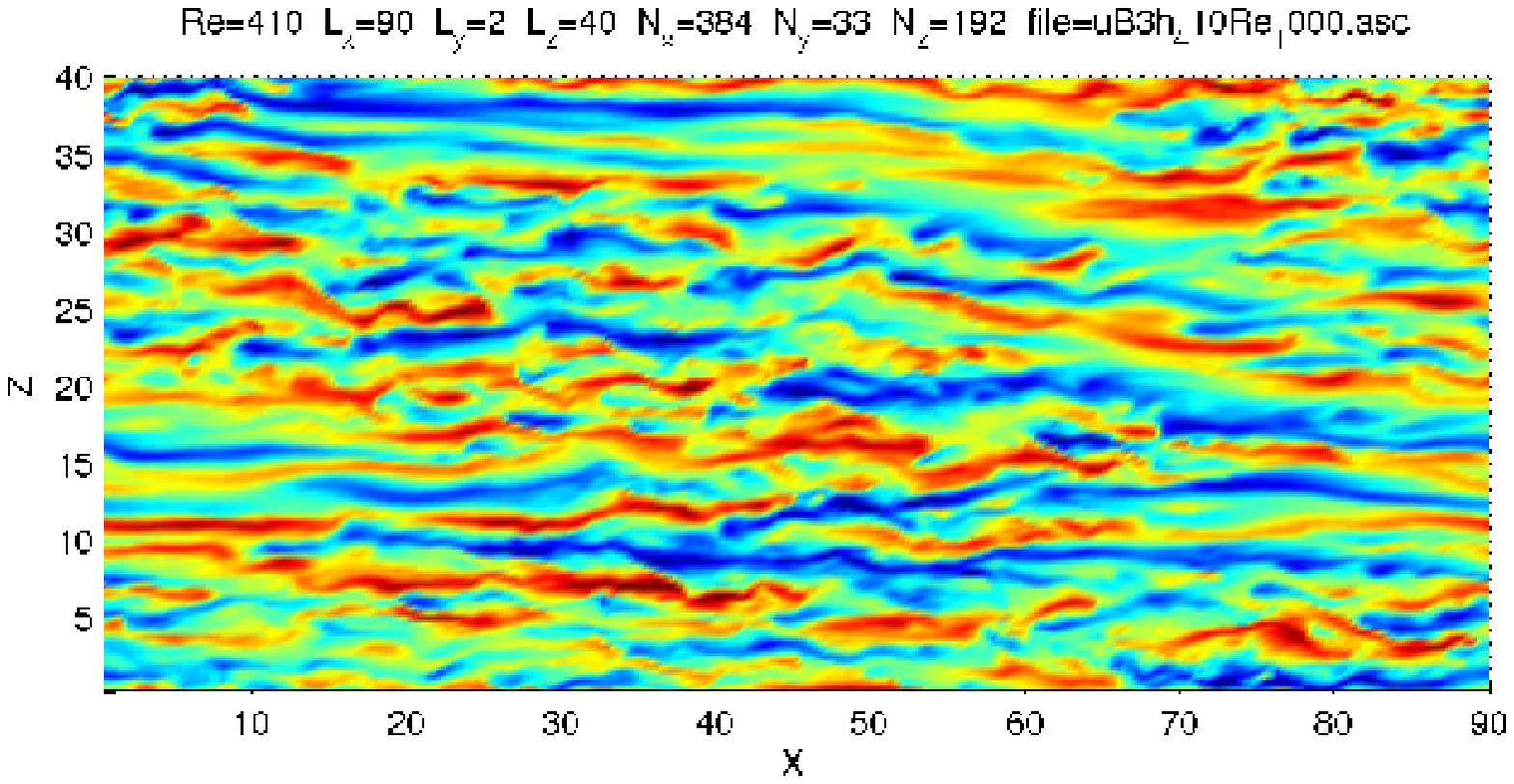}
\hspace{0.05\textwidth}
\includegraphics[width=0.28125\TW,clip]{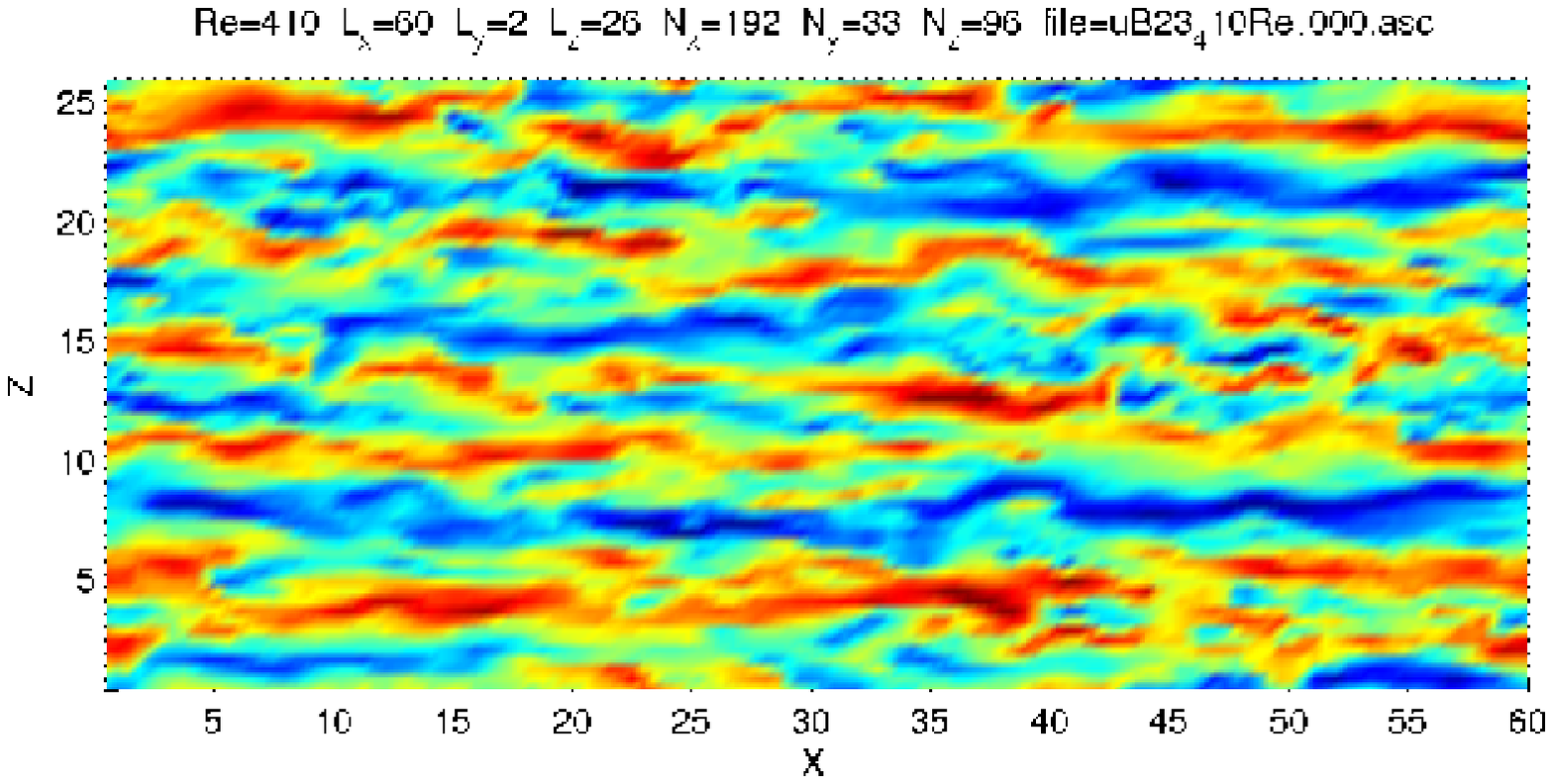}
\vspace{0.5cm}

\includegraphics[width=0.2461\TW,clip]{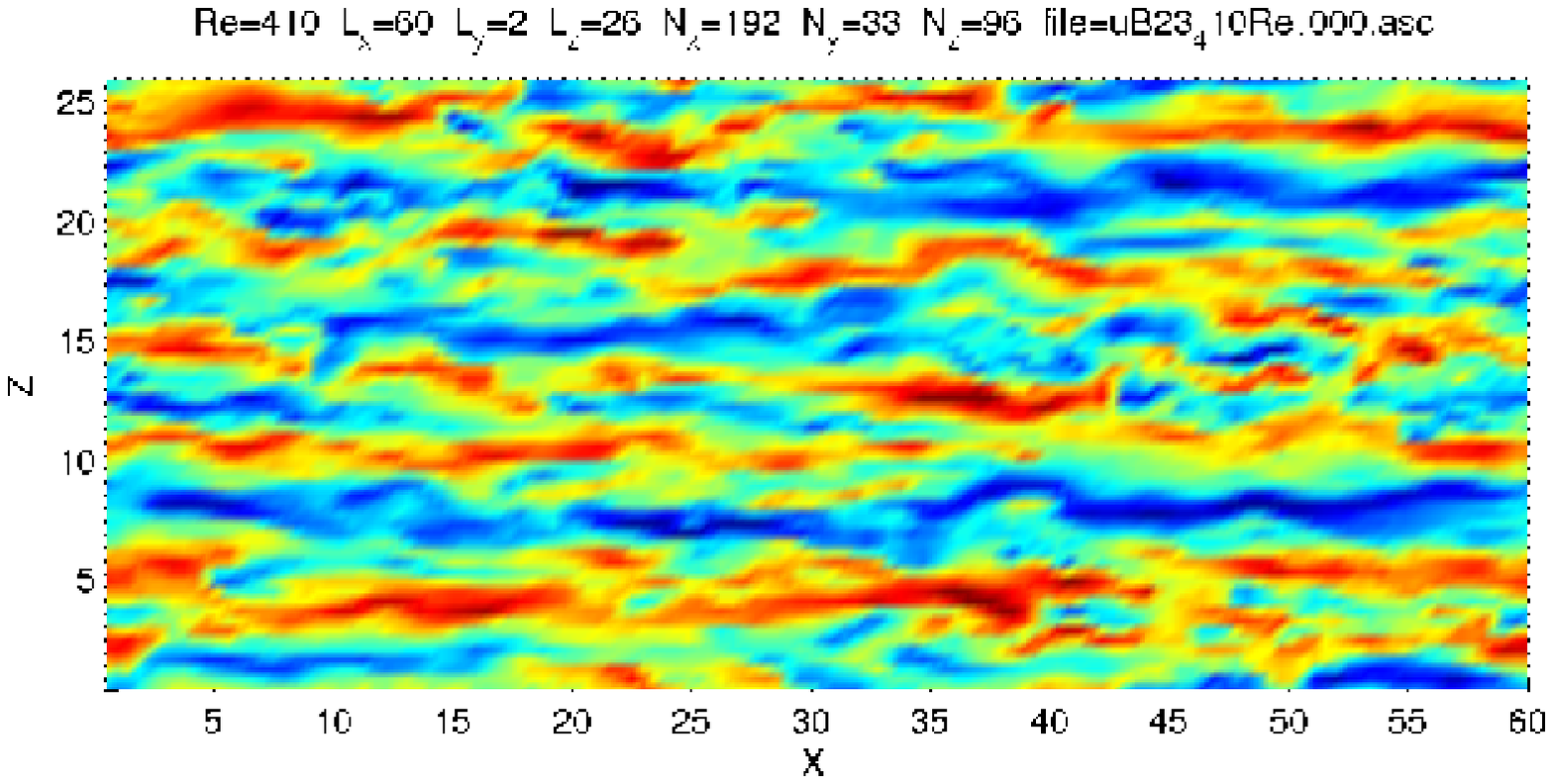}
\hspace{0.05\textwidth}
\includegraphics[width=0.2109375\TW,clip]{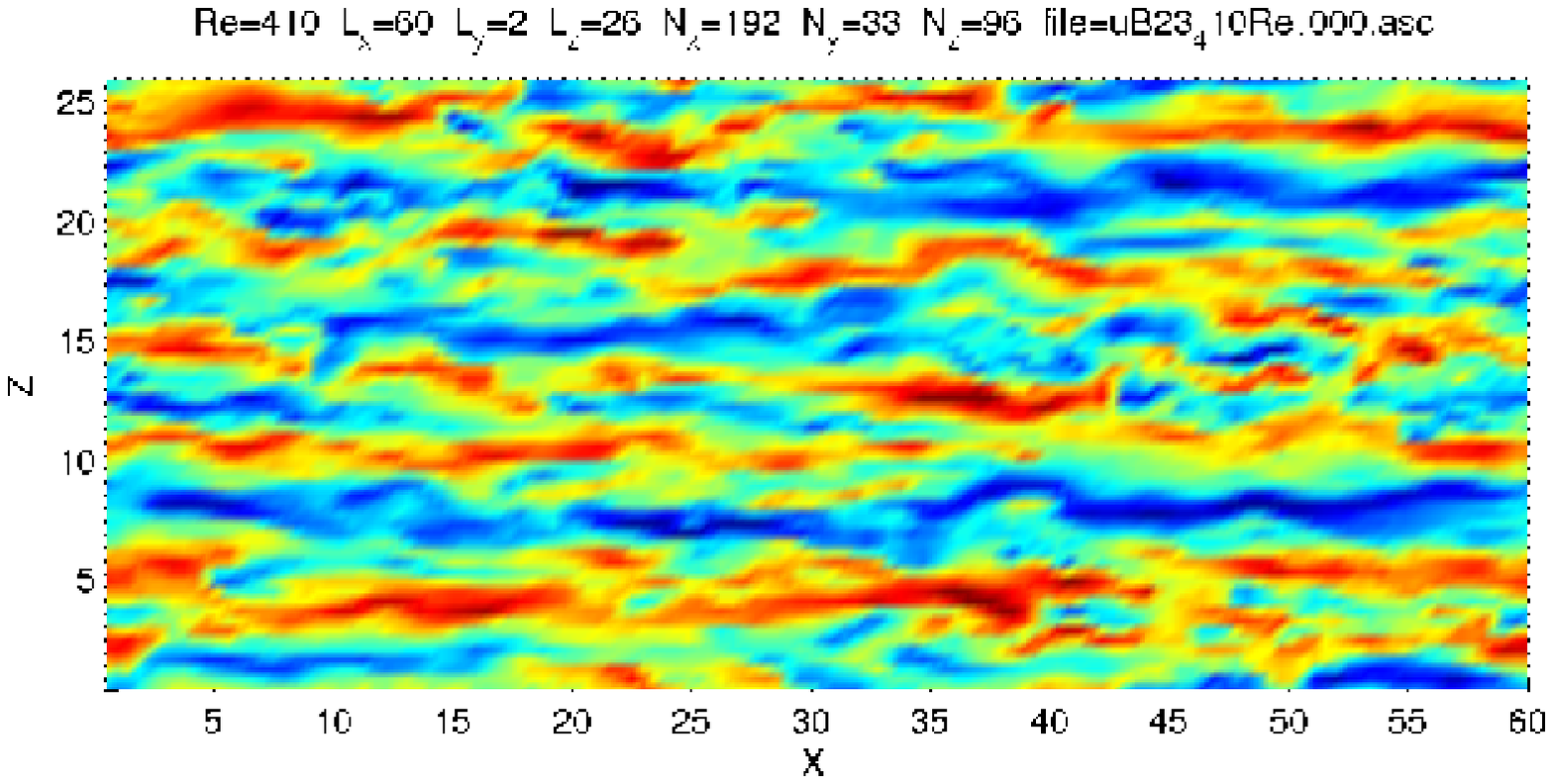}
\vspace{0.5cm}

\includegraphics[width=0.1125\TW,clip]{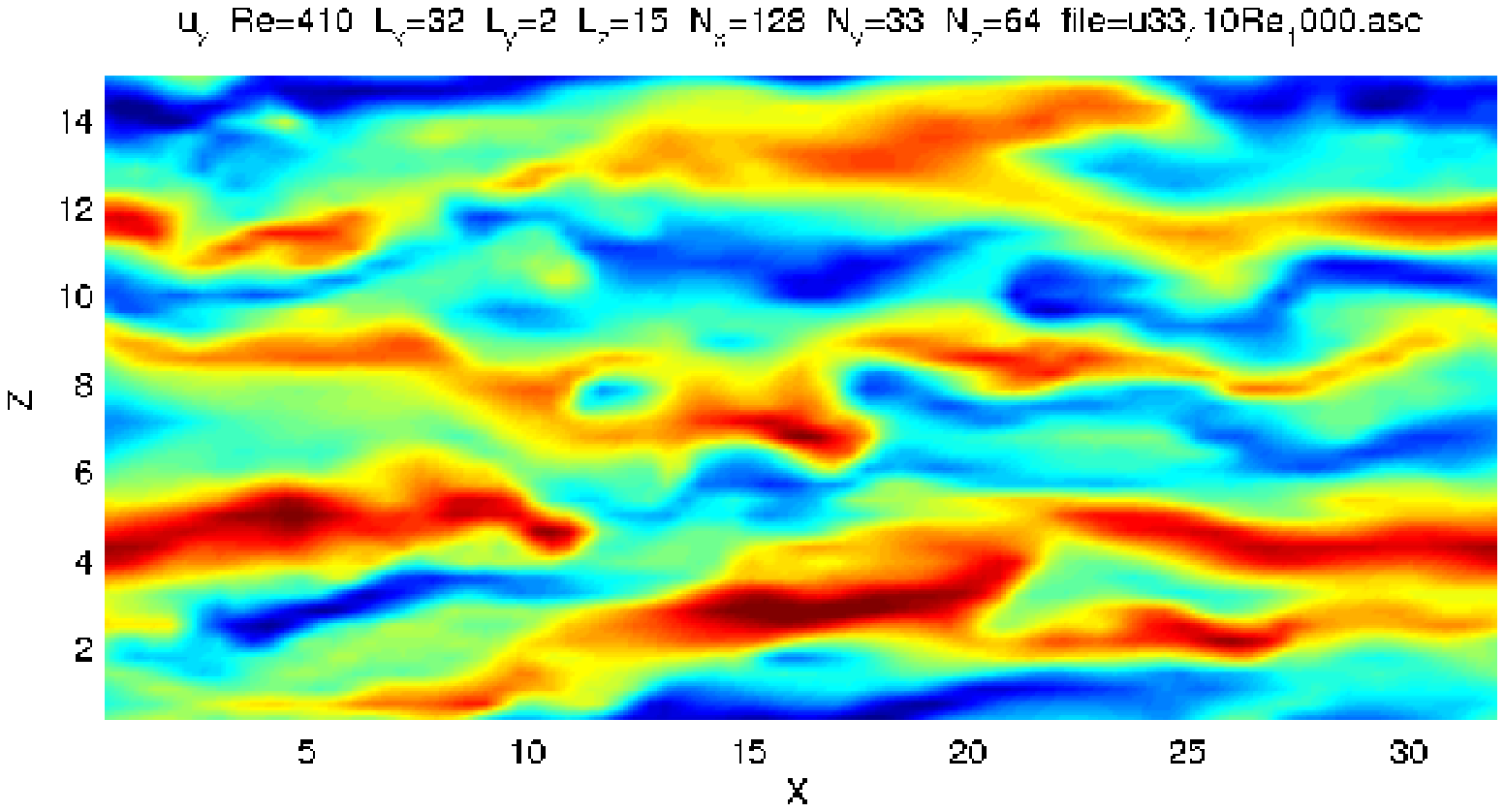}
\hspace{0.05\textwidth}
\includegraphics[width=0.084375\TW,clip]{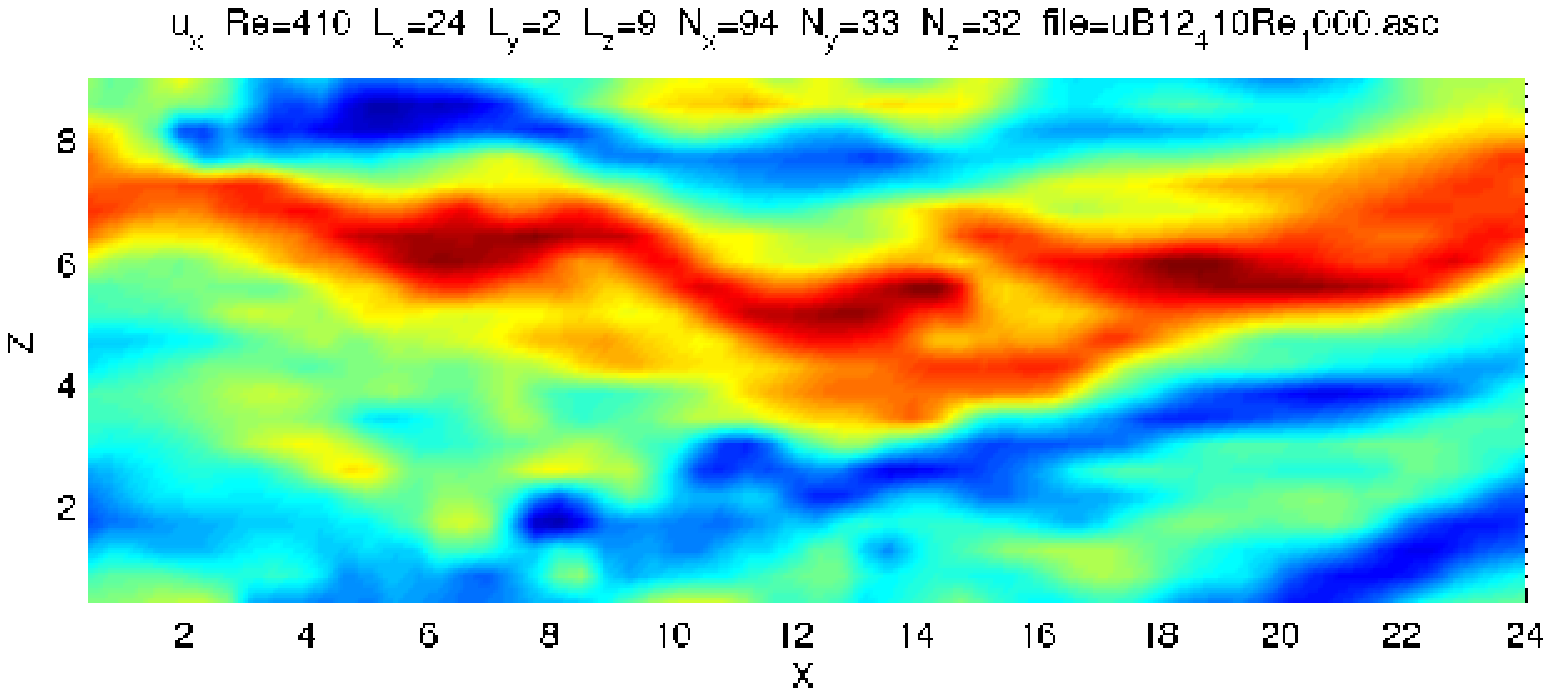}
\vspace{0.5cm}

\includegraphics[width=0.055195313\TW,clip]{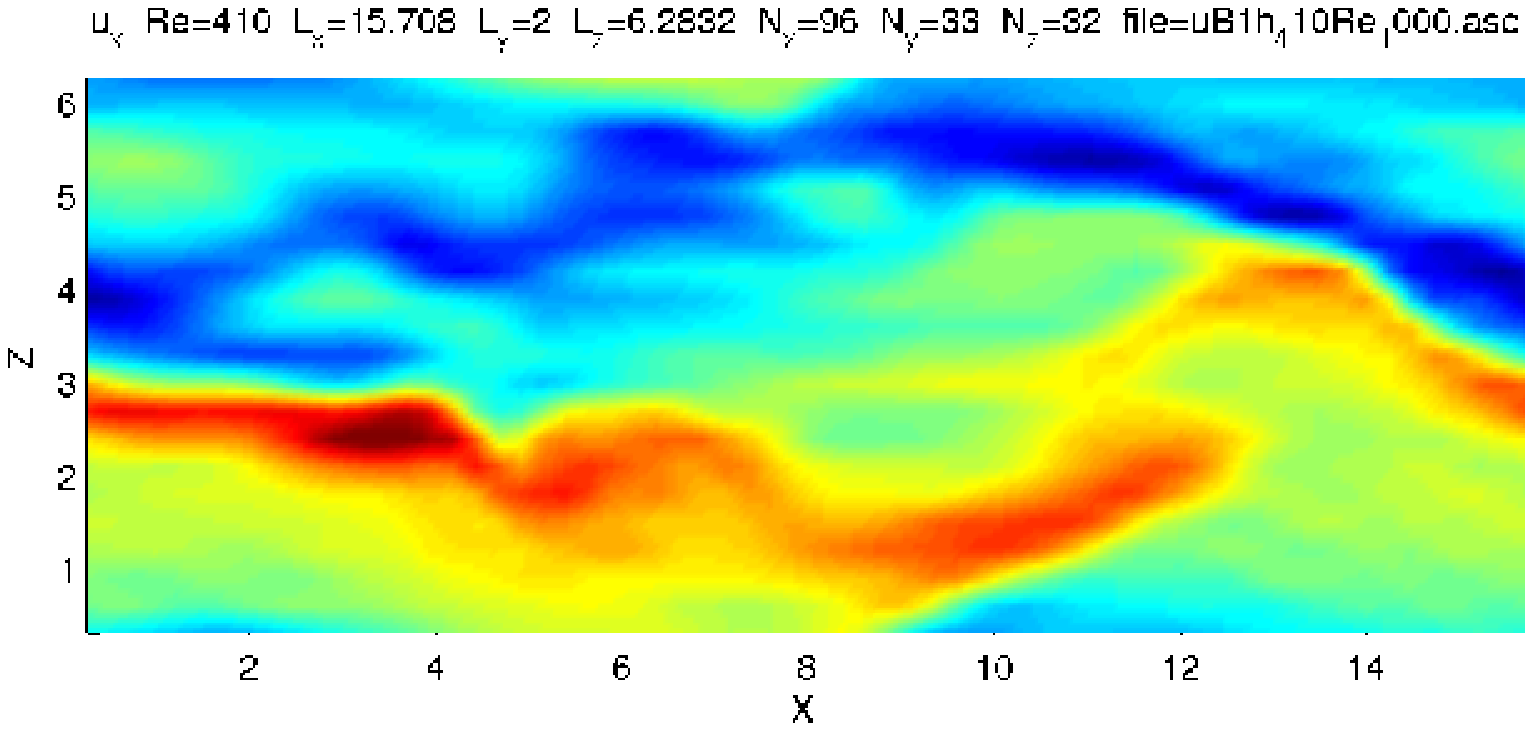}
\vspace{0.5cm}
\EC
\caption{(Color online) Streamwise velocity fluctuations ($u$) in the plane $y=0$ for different domains at $\RE=410$: $L=143.1$ (top-left), 109.6 (top-right), 98.5, 87.3, 76.2, 65.4, 35.3, 25.6, 16.9 (bottom).\label{fig3}}
\EF

The laminar base flow simply reads $\mathbf v_{\rm lam}=y\,\mathbf{\hat x}$, where $\mathbf{\hat x}$ is the unit vector in the streamwise direction.
The velocity perturbation $\mathbf{\tilde v}$ is obtained as $\mathbf{\tilde v}=\mathbf v - y\,\mathbf{\hat x}$, where  $\mathbf{v}:=u\,\mathbf{\hat x}+ v\,\mathbf{\hat y}
+w\,\mathbf{\hat z}$ is the instantaneous velocity field solution to the Navier--Stokes equations that are simulated. For further reference, Figure \ref{fig3} displays typical snapshots of the solutions obtained for $\RE=410$ in the different domains that we consider.
We illustrate the fluctuation field using $u$ evaluated in the plane $y=0$, which is a direct trace of the perturbation since $\mathbf v_{\rm lam}\equiv0$ there.
The different domains are displayed, starting with $L=143.1$ and decreasing the size from top-left to right-bottom.
The snapshots presented are approximately proportional to the domain sizes.
Colors red and blue correspond to $u\approx\pm0.5$.
These images show that the very large structures discovered by Komminaho {\it et al.} \cite{Kometal96}, in a domain of size $28\pi\times8\pi$ ($L\simeq91.5$)  at $\RE=750$ deep inside the featureless regime are already present at $\RE=410$, in the immediate vicinity of \RT.
 In all cases the flow fields are dominated by streamwise elongated (or streaky) structures which are more or less alternating in the spanwise direction.
 In the smallest domains, the development of these streaky structures is severely constrained.
 In larger domains, i.e. for $L=76.2$ and larger, pockets of laminar flow can be observed already at $\RE=410$. More details about these structures and their significance
as \RE\ is decreased will be discussed in the subsequent sections.

The results to be described now, in connection to the temporal/spatiotemporal issue in domains  of sizes varying from a few units of MFU to ones where patterns or bands appear, are all obtained using an `adiabatic protocol' to be described below according to which, starting from a turbulent state at high Reynolds number, \RE\  is reduced by steps of $\Delta\RE$ and the simulation is run for $\Delta T$, repeatedly down to the laminar regime.

\section {Results\label{3}}

\subsection {Fluctuations in varying domain size\label{3.1}}

Here we use two global measures of fluctuation intensity, the overall {\it rate of energy dissipation\/} per unit volume:
\BE \label{eqn_Diss}
\mathcal D := \frac{1}{L_x L_y L_z} \int_{0}^{L_x} \int_{-1}^{1}
\int_{0}^{L_z} \left(|\nabla u|^2 +|\nabla v|^2+|\nabla w|^2\right)
\; {\rm d}x \,{\rm d}y \,{\rm d}z\,,
\EE
where $L_y=2$, and the {\it rate of mechanical energy input}:
\BE \label{eqn_In}
\mathcal I := \frac{1}{L_xL_yL_z} \int_{0}^{L_x} \int_{0}^{L_z}
\left(\left.\frac{\p u}{\p y}\right|_{y=1}
+\left.\frac{\p u}{\p y}\right|_{y=-1}\right) \;{\rm d}x\,{\rm d}z\,.
\EE
They are normalized such that, for the laminar solution, both $\mathcal D$ and $\mathcal I$ are equal to 1.
For any flow field, on average $\mathcal D$ is equal to $\mathcal I$, and their instantaneous difference is a measure of the fluctuation of the system's energy:
\BE \label{eqn_E}
\mathcal E(t):=\frac{1}{L_x L_y L_z}\int_{0}^{L_x}\int_{-1}^{1}\int_{0}^{L_z}
\mbox{$\frac12$}
\| {\bf v}(x,y,z;t)\|^2 \;  {\rm d}x \,  {\rm d}y \,  {\rm d}z\,,
\EE
so that, directly from the Navier--Stokes equations, one derives
${\rm d}\mathcal E/{\rm d}t = \RE^{-1}(\mathcal I-\mathcal D)$.

\BF
\BC
\includegraphics[width=0.45\TW,clip]{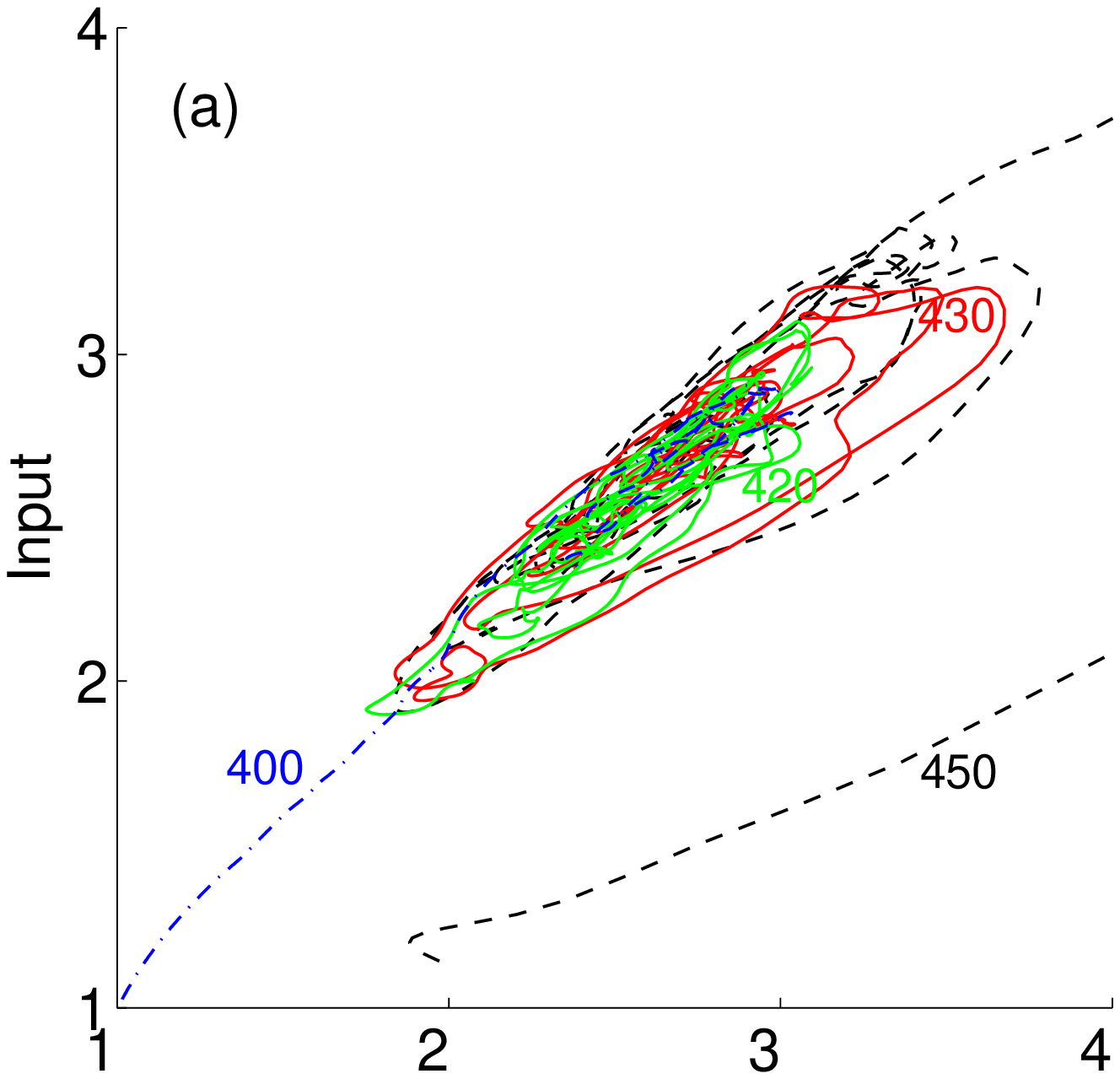}\hspace{2em}
\includegraphics[width=0.4223\TW,clip]{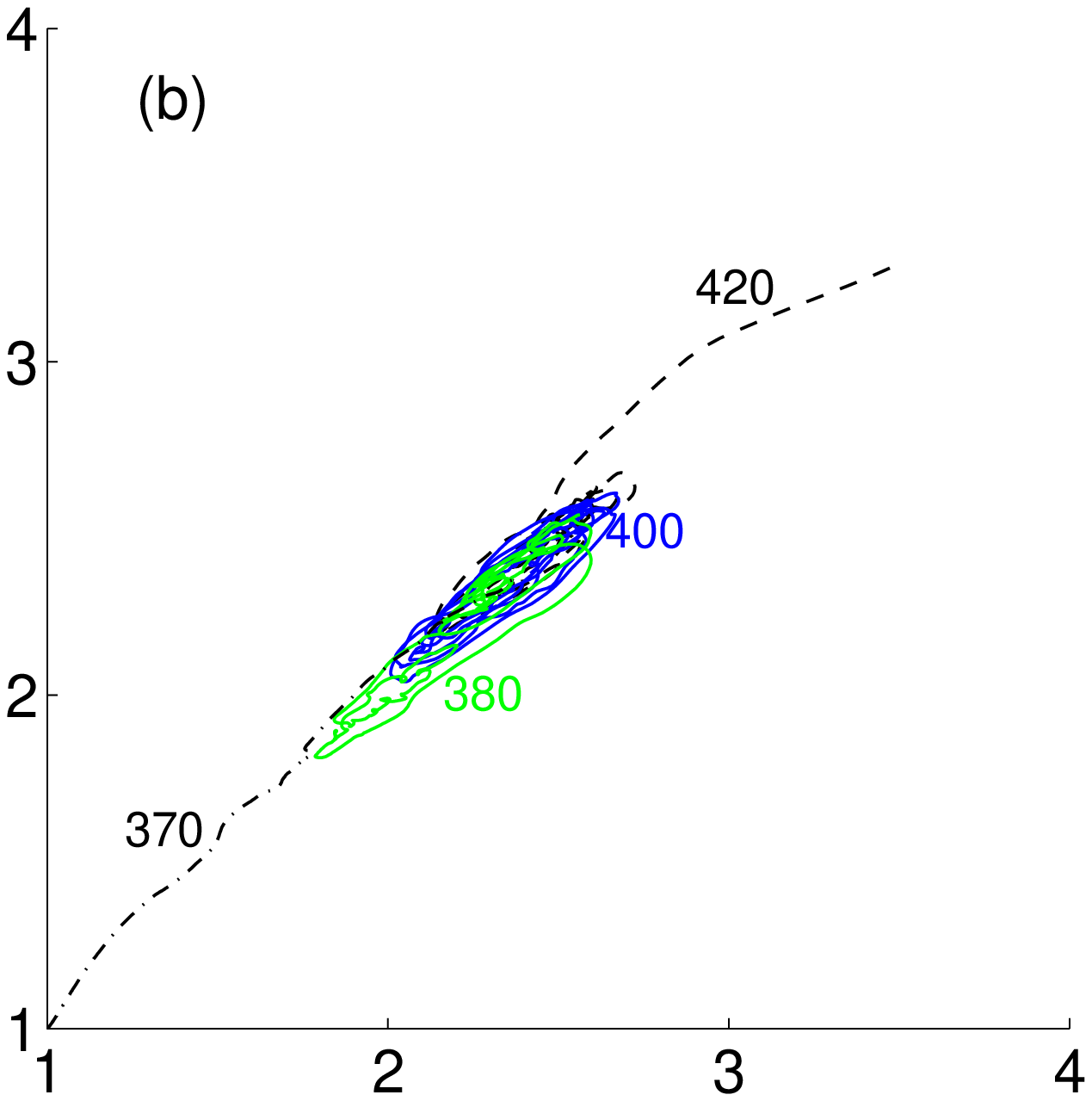}
\vspace{4ex}

\includegraphics[width=0.45\TW,clip]{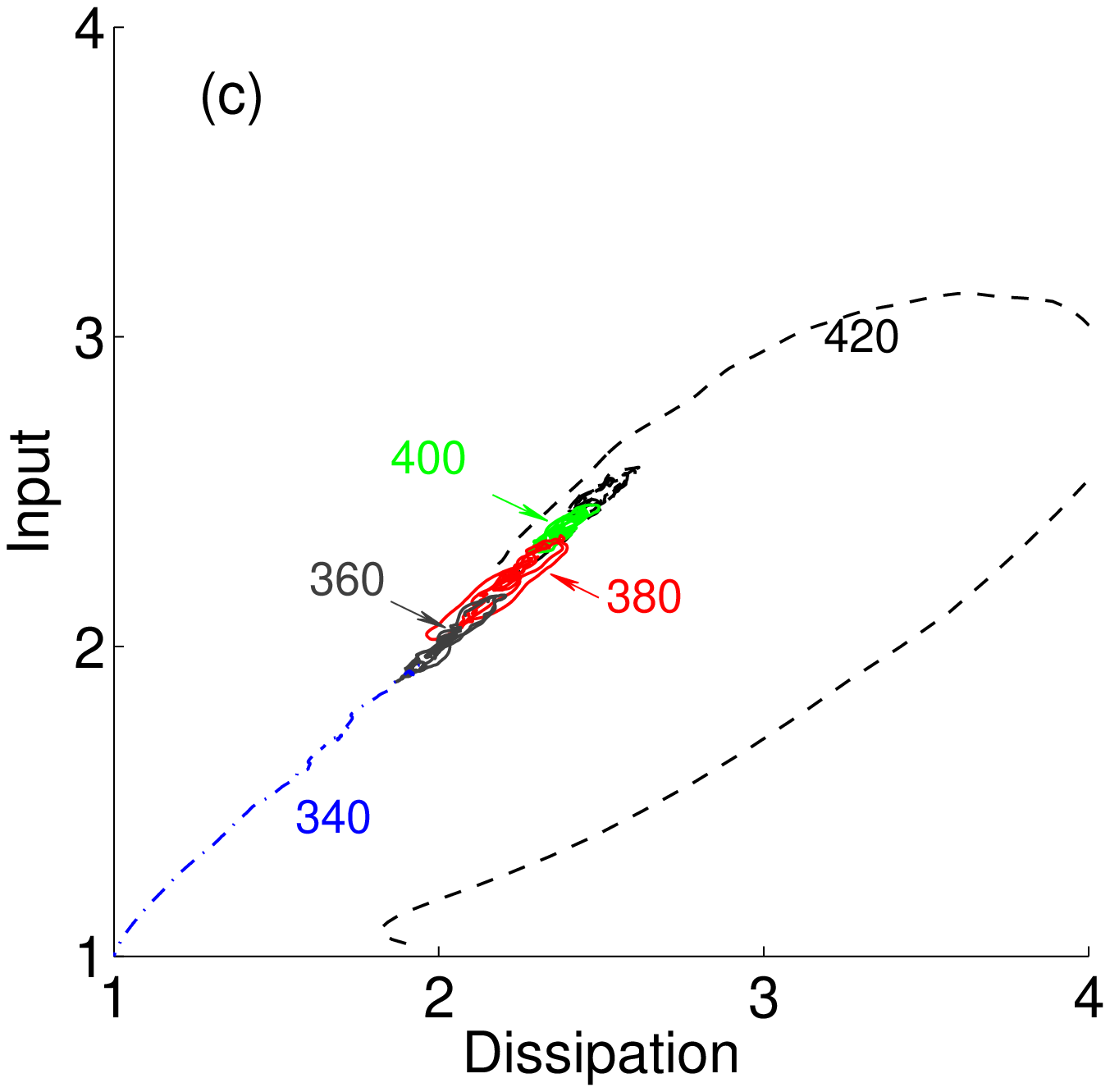}\hspace{2em}
\includegraphics[width=0.4223\TW,clip]{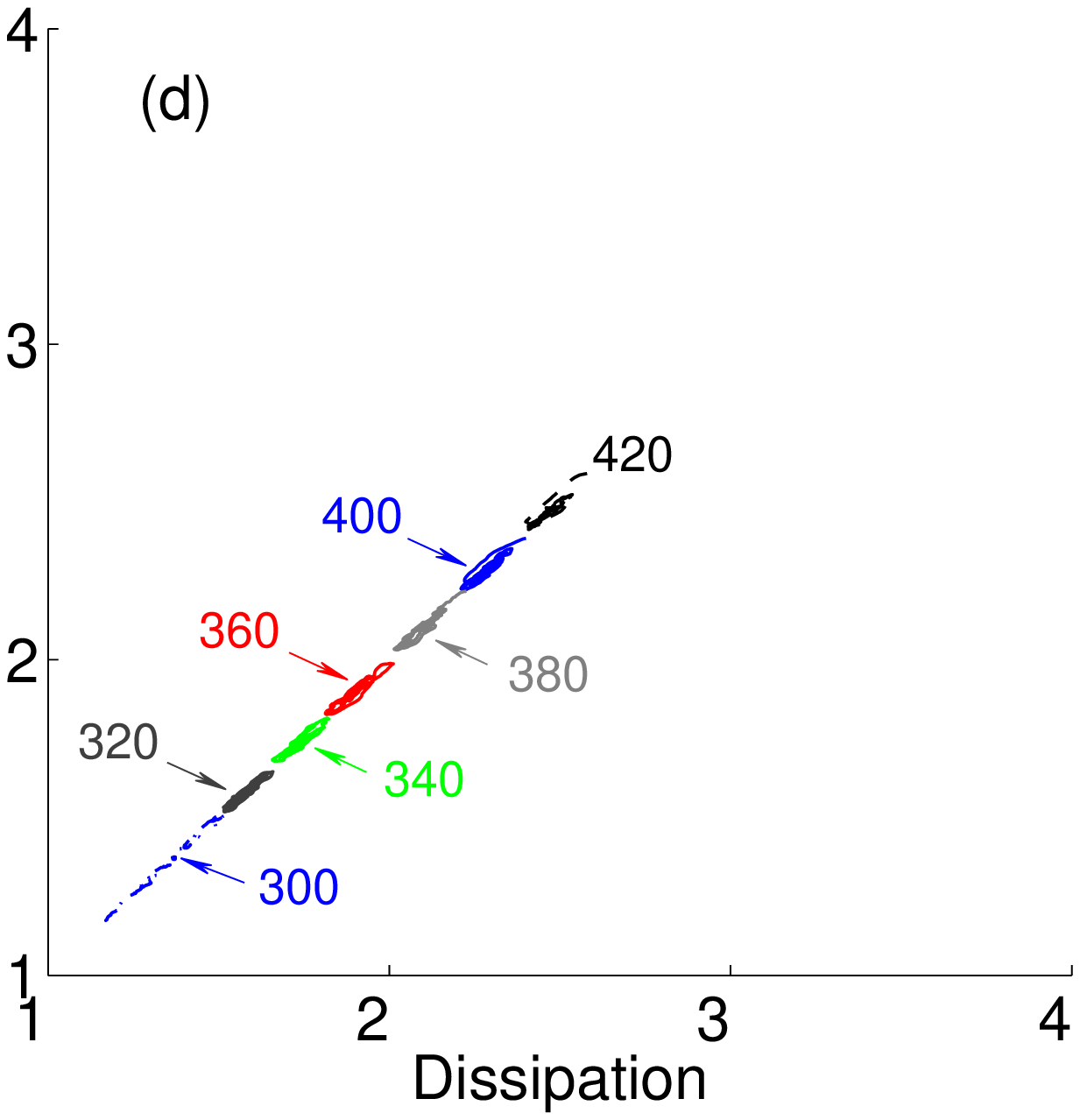}
\EC
\caption{(Color online) Input $\mathcal I$ {\it vs.} dissipation $\mathcal D$ as a
function of time at different \RE;
$L=16.9$ (a), $35.3$ (b), $76.2$ (c), and $143.1$ (d).
See text for details.\label{fig4}}
\EF

Figure \ref{fig4} shows the trajectories followed by the system as projected on the $\mathcal D$--$\mathcal I$ plane for four increasing domain sizes $L$: 16.9, 35.3, 76.2, and 143.1.
An adiabatic protocol is followed, similar to the annealing experiment performed by Schmiegel \& Eckhardt \cite{SchEck00} in a MFU-sized system or Barkley \& Tuckerman \cite{BarTuc05} in their oblique domain. We start at  $\RE=450$ for the smallest domain and at $\RE=420$ for the others.
By adiabatic, we mean that \RE\ is reduced by steps $\Delta\RE$ every $\Delta T$, the final state at a given \RE\ being used as initial condition for the simulation at $\RE-\Delta\RE$. Usually we take $\Delta T=10^3$ and $\Delta\RE=10$ for $\RE\ge320$ and  $\Delta\RE=5$ below 320.
Not all the traces are shown in order not to over-print the figure.
Traces tending to the (1,1) point in the $\mathcal D$--$\mathcal I$ plane express `turbulent$\>\to\>$laminar' breakdown, which is the case for $\RE=400$ when $L=16.9$, $\RE=370$ for $L=35.3$, etc.
This is also the case for $\RE=300$ when $L=143.1$ but the simulation has been interrupted before complete relaxation toward the laminar regime.
For the smallest domain, it is observed that the trajectory quickly falls in the neighborhood of the diagonal $(\mathcal D=\mathcal I)$ and then continues to revolve in a region along this line.
Decreasing \RE\ results in a slight decrease in the wandering but the point representing the state of the system stays along the diagonal and close to it.
As the domain size is increased, the amplitudes of the excursions along the diagonal, and away from it, decrease and the dynamics seems slower;
the transition from one \RE\ to another also seems smoother.
These characteristics can be understood as the result of averaging over larger domains, in connection with the extensive character of the featureless turbulent regime examined in \S\ref{3.2} below.

A complementary piece of information is obtained from the {\it distance to laminar flow\/}, whose time average was proposed in Fig.~\ref{fig2} to characterize the flow regime on a global scale.
As a relevant measure of the distance, we take the volume-averaged root-mean-square
value of the velocity perturbation $\mathbf{\tilde v}$, here denoted $\Urms$.
Time series of this quantity during simulations performed according to the adiabatic protocol described earlier are shown in Figs.~\ref{fig5} and~\ref{fig6} in two representative cases, $L=16.9$ and $L=143.1$, respectively:

$\bullet$\quad As seen in Fig.~\ref{fig5} for $L=16.9$, while the average of \Urms\ for different \RE\ does not vary much, at some reduced \RE\ the flow suddenly relaminarizes.
\BF
\includegraphics[width=\TW,clip]{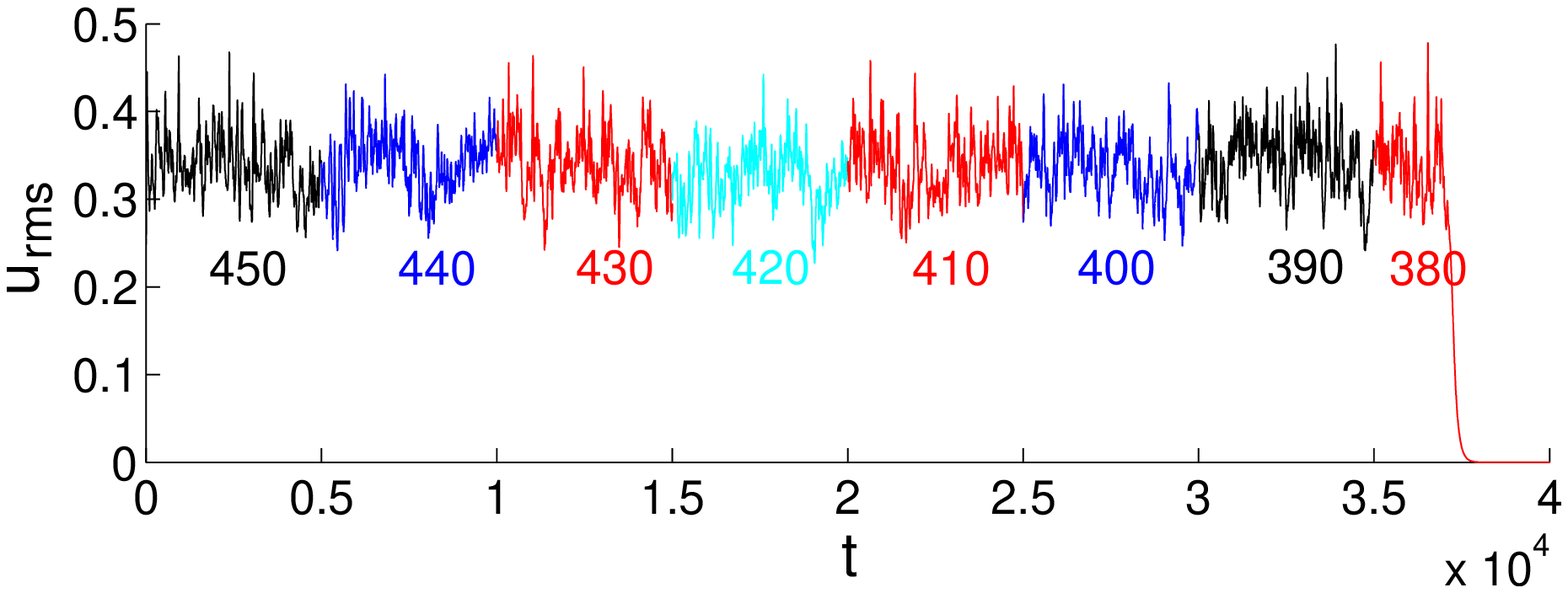}
\caption{(Color online) Distance to laminar flow (\Urms) at different \RE\ for $L=16.9$. Adiabatic protocol with time series of duration $\Delta T=5\cdot10^3$.
\label{fig5}}
\EF
This process was found to be probabilistic, e.g. by Schneider {\it et al.} \cite{Sch_etal10}, and explained as arising from chaotic transients associated to the dynamics around a tangle in phase space. Snapshots taken during the decay of the chaotic state at
$\RE=380$ are displayed in Fig.~\ref{fig5b}, bottom.
\BF
\includegraphics[width=\TW,clip]{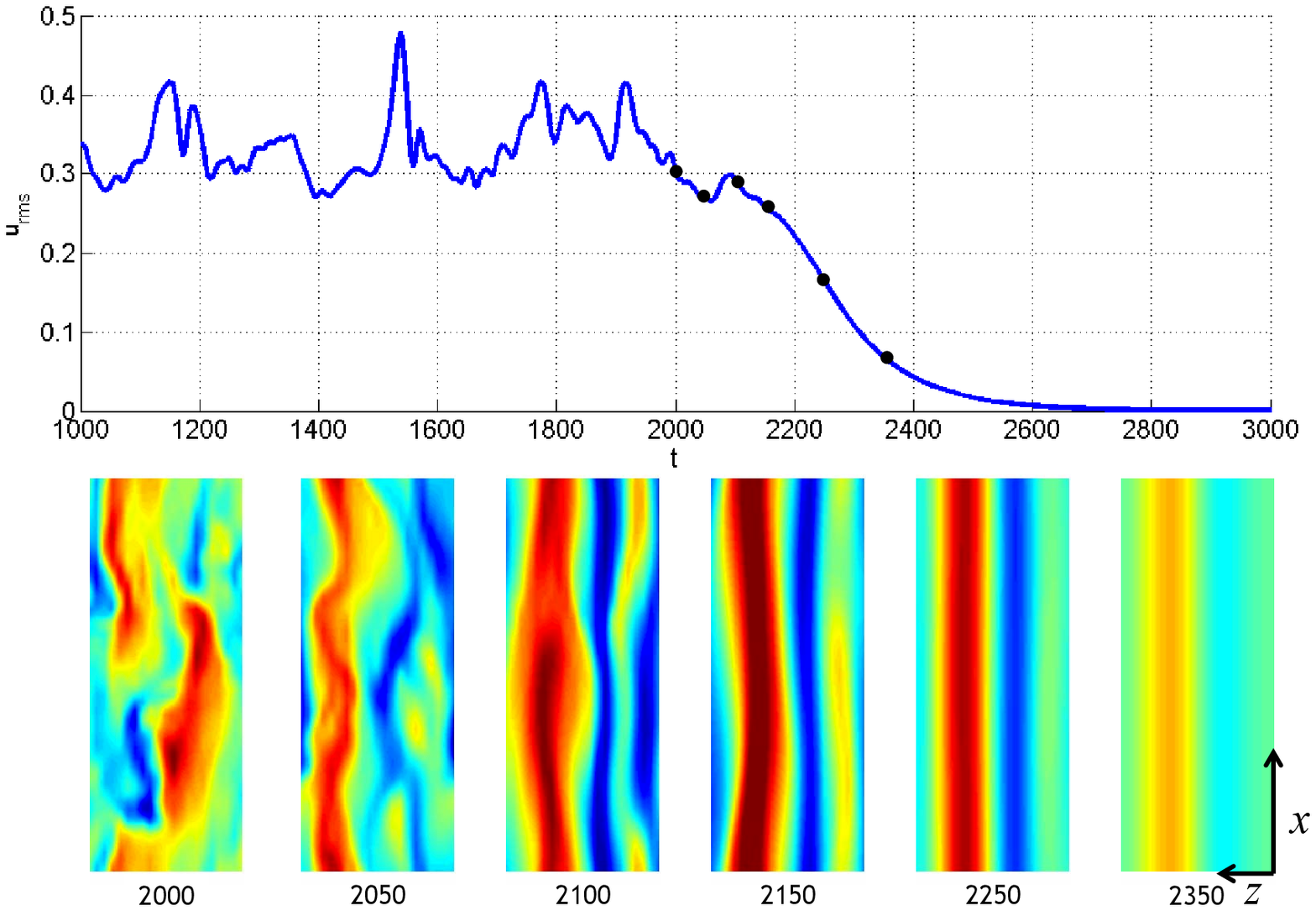}
\caption{(Color online) Top: Distance to laminar flow (\Urms) during the end of the transient at $\RE=380$ for $L=16.9$.
Bottom: Snapshots during decay; time, corresponding to points on the graph, is indicated below the images; exceptionally, the streamwise direction is along the vertical.
\label{fig5b}}
\EF
Whereas images taken before decay look similar to the one at $t=2000$, breakdown of the chaotic state is seen as a fast damping of small scale structures ($t=2050$), leaving just a pair of streaks ($t=2100$, $2150$) that progressively fade away ($t=2250$, $2350$).
Here decay is observed for $\RE=380$ whereas a similar decay happened at $\RE=400$ in the experiment reported in Fig.~\ref{fig4}a, with different initial conditions and a different protocol ($\Delta T=10^3$ instead of $5\cdot10^3$ here) but this merely expresses the probabilistic character of chaotic transients whose lifetimes are strongly sensitive to initial conditions \cite{EcFaScSc08}.
It should be noted that the exact size of the domain certainly matters for such small systems since, using their annealing protocol at a velocity corresponding to the rate $\Delta R/\Delta T =2\times10^{-3}$ of Fig.~\ref{fig5}, Schmiegel \& Eckhardt \cite{SchEck00} found that transient chaotic dynamics with lifetime of order 2500 time units could be maintained down to $R\simeq310$ in a system of size $2\pi\times\pi$, whereas in a system of size $5\pi\times2\pi$  we obtain short-lived chaotic transients already at $R=380$ (though we have not performed a detailed statistical study). On the other hand, a definite mark of spatiotemporal dynamics will bear a much weaker sensitivity to the precise in-plane dimensions of the system.

$\bullet$\quad When $L=143.1$, upon further lowering of \RE\ below 400, the average value of \Urms\ is seen to decrease regularly with \RE\ until relaminarization occurs, here for $\RE=300$. Also shown in Fig.~\ref{fig6}
\BF
\centerline{\includegraphics[width=\TW,clip]{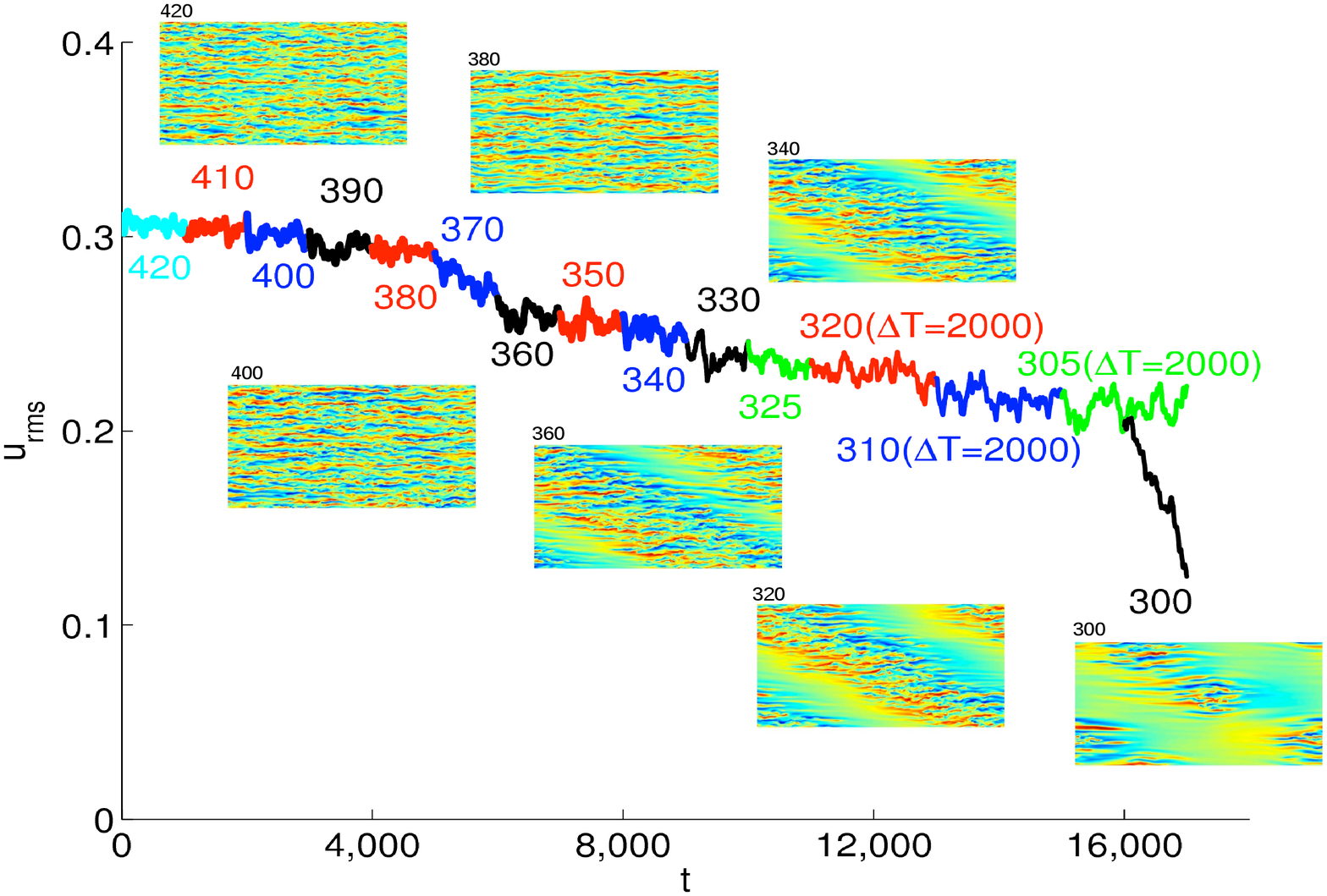}}
\caption{(Color online) Distance to laminar flow (\Urms) at different \RE\ for $L=143.1$ ($\theta=26.6^\circ$).
Inserted figures show snapshots of $u$ at various \RE\ (indicated on top-left) after simulation over $\Delta T=10^3$.
\label{fig6}}
\EF
are the snapshots of $u$ at various \RE\ (indicated on top-left) at the end of a simulation step ($\Delta T=1000$).
At $\RE=380$, {\it laminar troughs\/}, i.e. small patches where turbulence seems depleted can be observed (and not yet patterns).
Upon further reduction of \RE\ the troughs join to form a well defined pattern of oblique bands, alternatively laminar and turbulent (snapshot at $\RE=360$).
The subsequent decrease in \Urms\ is due to the increase in the width of the laminar band.
The final decay to laminar flow is neither sudden nor uniform in space as in small domains, but, through a local breaking of the band leaving separated turbulent patches (see snapshot at $\RE=300$) that recede and disappear.
Only at the very last stage, when the turbulent patches have reached a small size, they do collapse under the effects of viscous dissipation like the flow in small domains.
All along the decay, the internal structure of the turbulent spots is not fundamentally different from that of spots growing from a finite-amplitude localized disturbance at higher Reynolds numbers, as obtained experimentally \cite{TiAl92,Bo98,Betal98} or numerically \cite{LunJoh91,DugSchDan10}.
It should be noted that, here, decay is observed at $\RE=300$ while experiments at large aspect ratio would rather suggest decay for $\RE<\RG\approx 325$.
A first reason for this delayed breakdown may be that periodic boundary conditions tend to stabilize the pattern.
A second, certainly more important, reason is that according to our adiabatic protocol, \RE\ is changed by $\Delta\RE$ every $\Delta T$ and that $\Delta\RE$ may be too large and $\Delta T$ too small.
Basically, the systems `knows' at which value of \RE\ the simulation is running only after several viscous relaxation times $\tau_{\rm v}=\RE\, \tau_{\rm a}$, where $\tau_{\rm a}$ is the
advection time $h/\UP$, which is also our time unit.
Accordingly, $\Delta T=10^3 \sim 3 \tau_{\rm v}$, which is barely sufficient to reach the steady state.
This is all the more true that, owing to the subcritical character of the laminar--turbulent coexistence, we should allow for long waiting times implied by the nucleation of the stable (laminar) state within the metastable (turbulent) state at the origin of the breaking of the band that causes the decay \cite{Mazz}.
The need for long lasting simulations in wide enough systems was indeed a strong motivation to considering under-resolved DNS as a modeling strategy \cite{MaRo10}.

\subsection{From temporal to spatiotemporal dynamics through probability distributions\label{3.2}} 

From a statistical point of view, energy fluctuations are best characterized by their probability distribution functions (PDFs).
Figure \ref{fig7}
\BF
\BC
\includegraphics[width=0.48\textwidth]{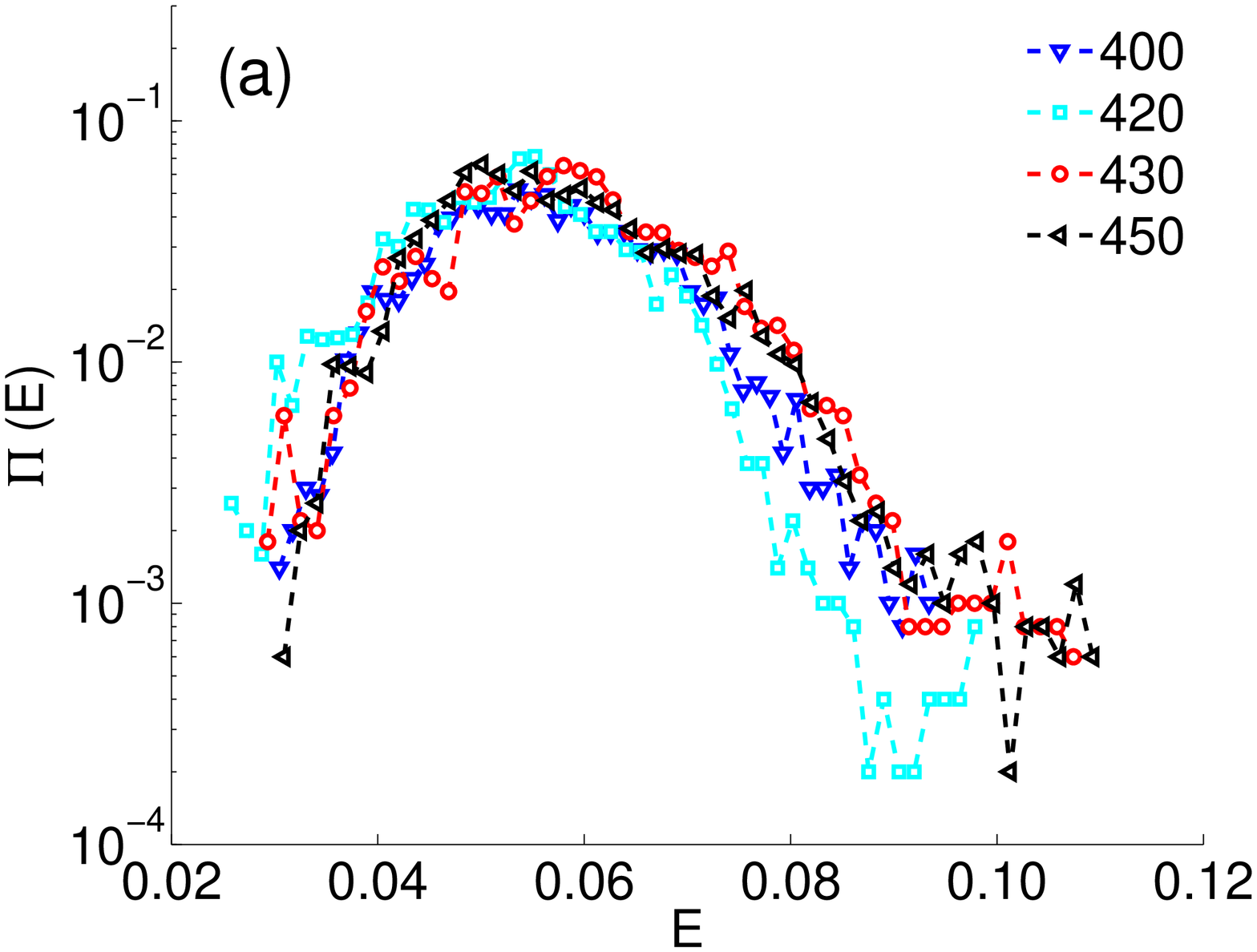}
\includegraphics[width=0.48\textwidth]{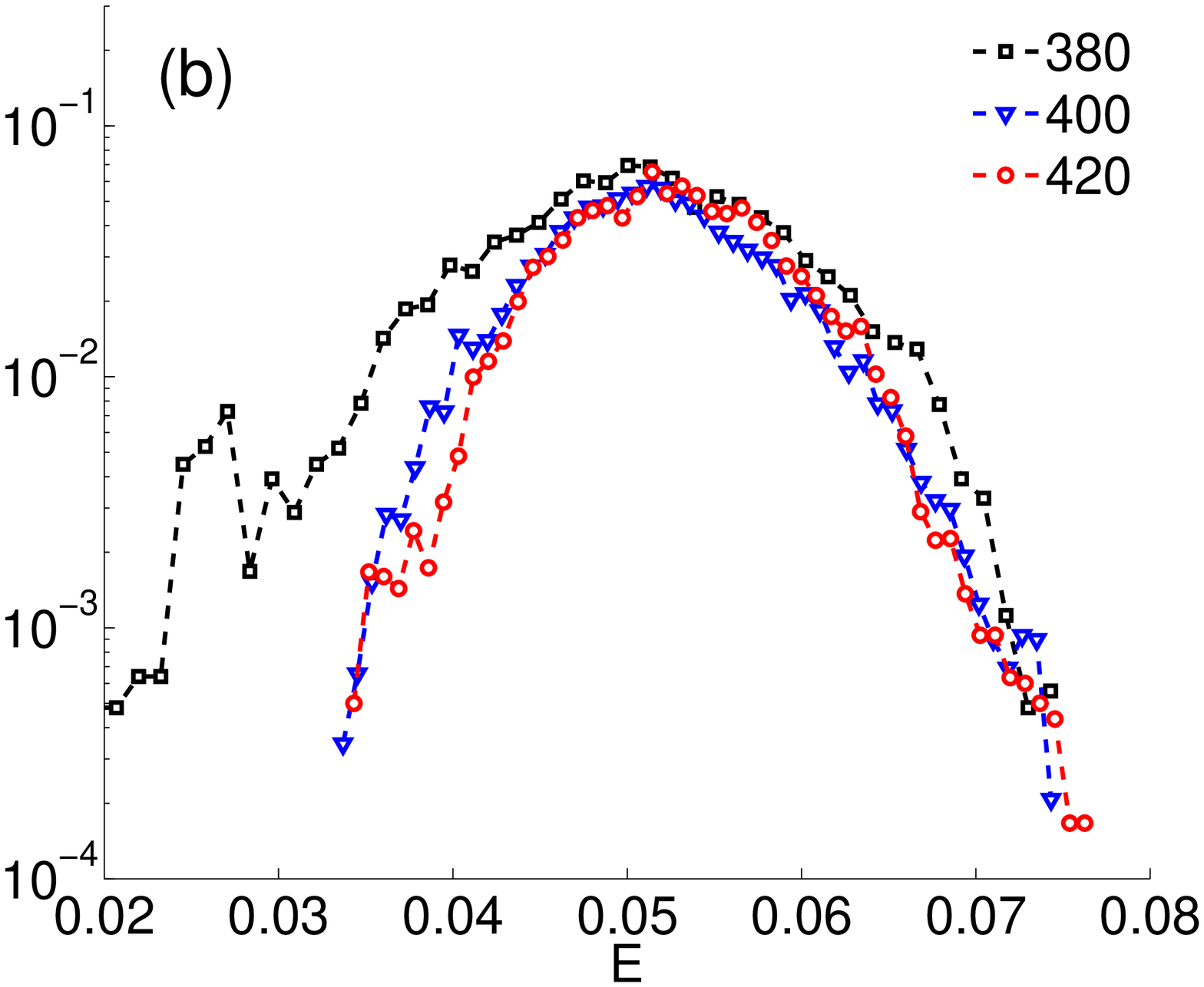}

\includegraphics[width=0.48\textwidth]{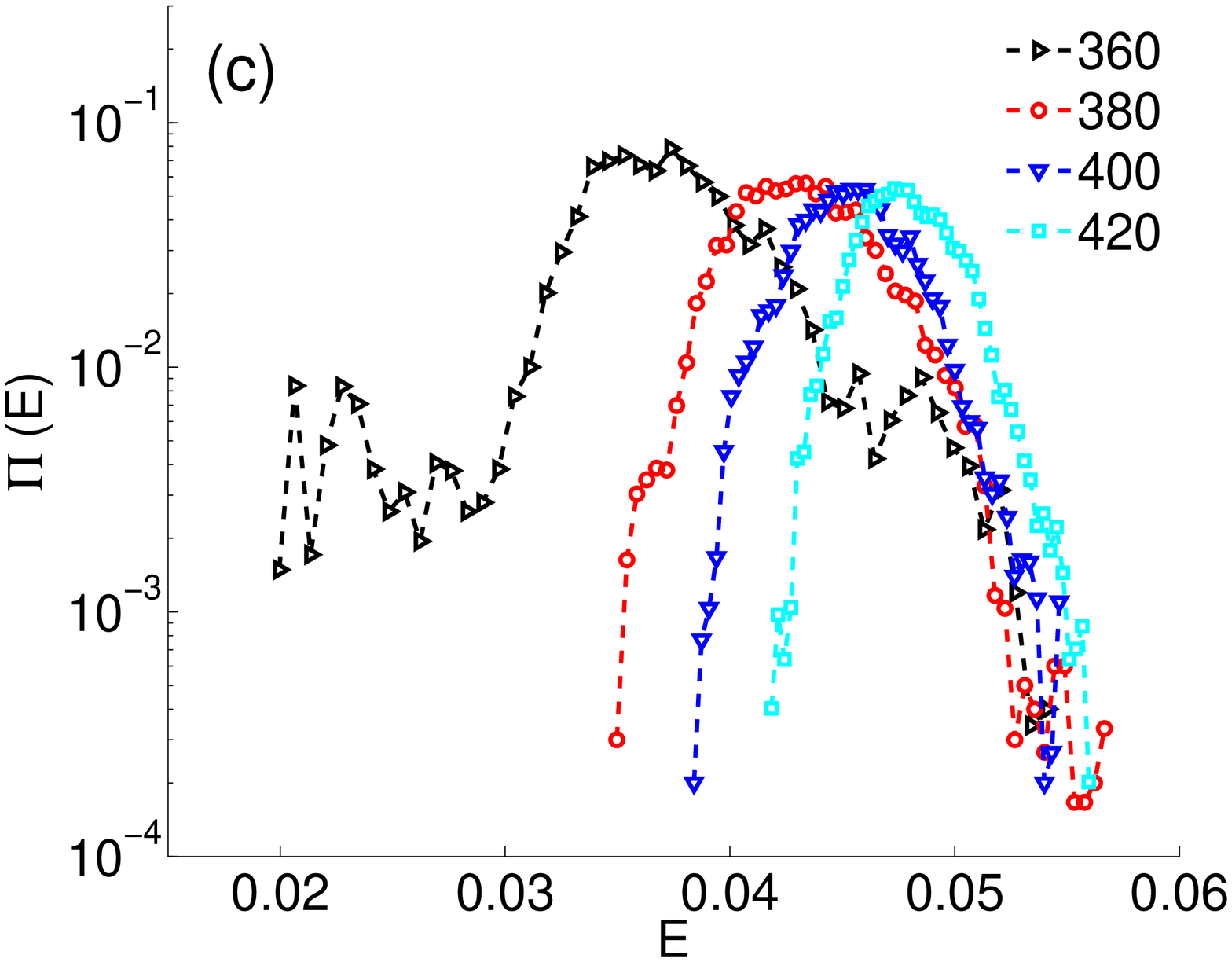}
\includegraphics[width=0.48\textwidth]{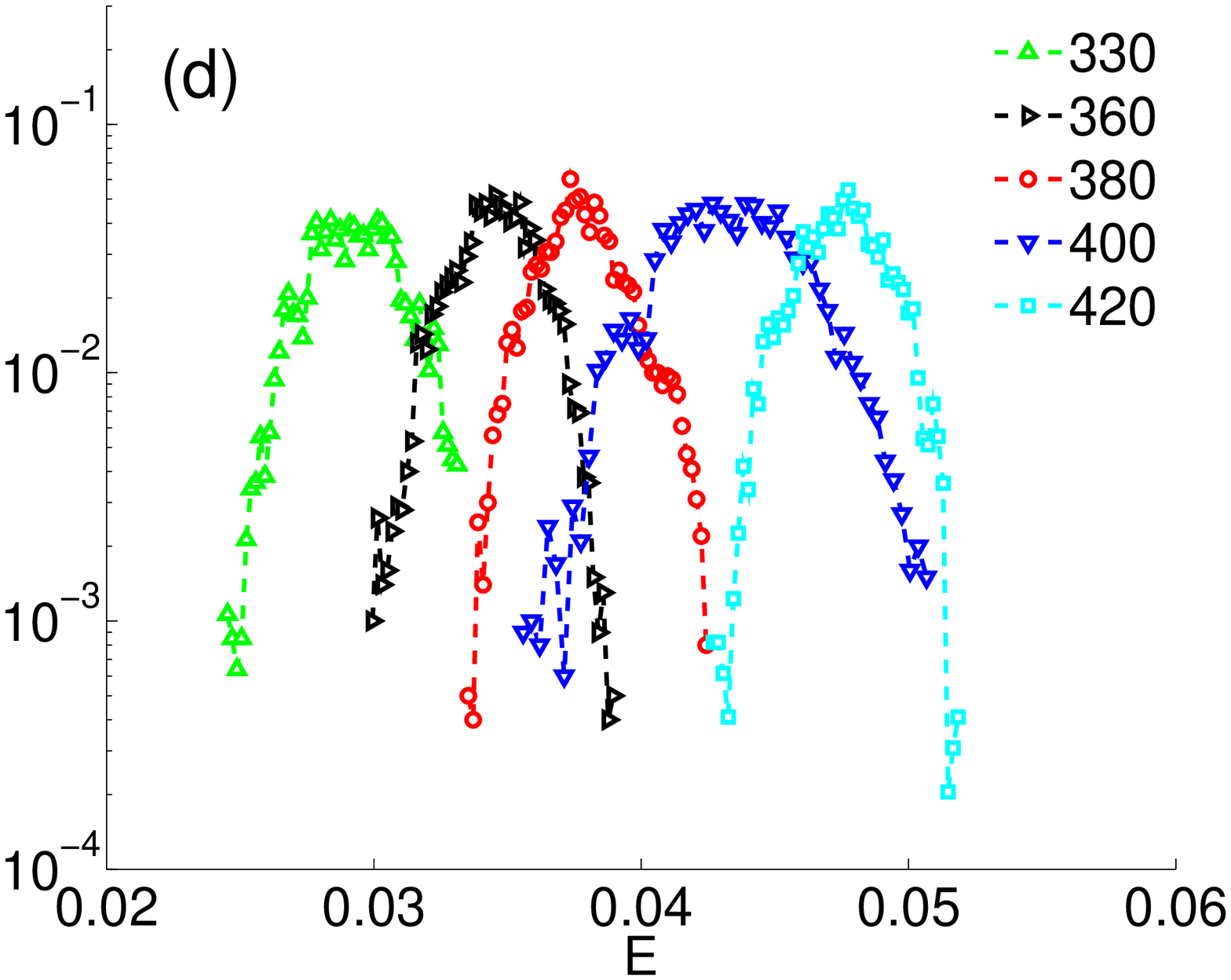}
\EC
\caption{(Color online) Probability distribution functions $\Pi(E)$ as functions of $E$ for different system sizes; values of \RE\ are indicated in the legends. (a) $L=16.9$; (b) $L=35.3$; (c) $L=76.2$; (d) $L=98.5$. Notice the different scales for $E$ on the horizontal axis.\label{fig7}}
\EF
plots, in lin-log coordinates, the normalized histograms $\Pi(E)$ of $E(t)$ defined as the volume average of $\frac12 \mathbf{\tilde v}^2$, i.e. $E=\frac12 (\Urms)^2$ and recorded for four different domains, $L=16.9$, 35.3, 76.2, and 98.5 at different values of \RE, during runs of duration up to $3\cdot10^4$ time units.
This is much longer than for Fig.~\ref{fig5} or \ref{fig6} and sufficient for the present purpose since the remark about the duration of the simulation does not apply as long as the system is not on the verge of decaying.
All the curves display a marked hump with some variations.
For the smallest domain,  Fig.\ref{fig7}a, both the mean value (see also Fig.~\ref{fig5}) and the most probable value (MPV) of $E(t)$ do not change significantly with \RE;
the tail present at large $E$ is presumably a signature of the underlying chaotic
dynamics.
For $L=35.3$ in Fig.~\ref{fig7}b the curves are roughly parabolic (which corresponds to Gaussian distributions), except for  $\RE=380$. The MPV stays fixed with \RE\ like for  $L=16.9$ whereas the extended tail towards smaller $E$ for $\RE=380$ suggests approaches to the laminar state which, when sufficiently marked, lead to turbulence breakdown, so that the system likely not far above the value below which decay can happen in a short time.
The same tendency is observed in Fig.~\ref{fig7}c for an larger domain, $L=76.2$, but now a downward shift of the MPV of $E$ is observed, which is best attributed to the existence and growth of the laminar fraction as \RE\ is decreased, laminar fraction that will be more conspicuous at larger $L$ as shown in Fig.~\ref{fig6}.
The PDF also displays a low-end tail for $\RE=360$ which again implies that turbulence breakdown can take place if one is patient enough in pursuing the simulation, and will do so more easily at a lower \RE.
The last case is for $L=98.5$ in Fig.~\ref{fig7}d, showing cleaner parabolic shapes and the systematic shift in the MPV of $E$.
Since $\RE=330$ is sufficiently above $\RG=325$, the system is not at risk to decay and the PDF has no low-$E$ tail.

From these curves, one main feature emerges: the position of the MPV may be variable or not.
Snapshots of the solutions show that the decrease  of the MPV with \RE\ reflects the presence of a larger and larger region of the system which has returned to laminar flow, temporarily or persistently, and the PDFs tell us that it is in a statistically significant way.
We interpret this as the signature of the change in the dynamics from temporal to spatiotemporal and conclude that one can decide this issue blindly from the consideration of time series of $E$ only.
From the variation of the MPVs, one can locate the change for $L$ around 76.2.
A secondary feature is the presence or not of low-end tail in the PDF of $E$, which is an indication of the robustness of the state considered.
In the absence of a tail, the distribution is essentially Gaussian and the system is not at a risk to decay at the given value of \RE.
If an exponential tail is present, it means that the decay probability is small but significant, which means that if the experiment is long enough, the flow will decay.

Let us now consider the featureless turbulent regime at $\RE=420$ where the PDFs displayed in Fig.~\ref{fig7} are all nicely one-humped.
A quantitative characterization of their shapes is obtained by fitting $\ln\Pi(E)$ against a polynomial in the form $aE^2 + bE + c$.
The width of the distribution can be defined from $a$ as $\sigma:=|a|^{-1/2}$.
\BF
\BC
\includegraphics[width=0.8\TW]{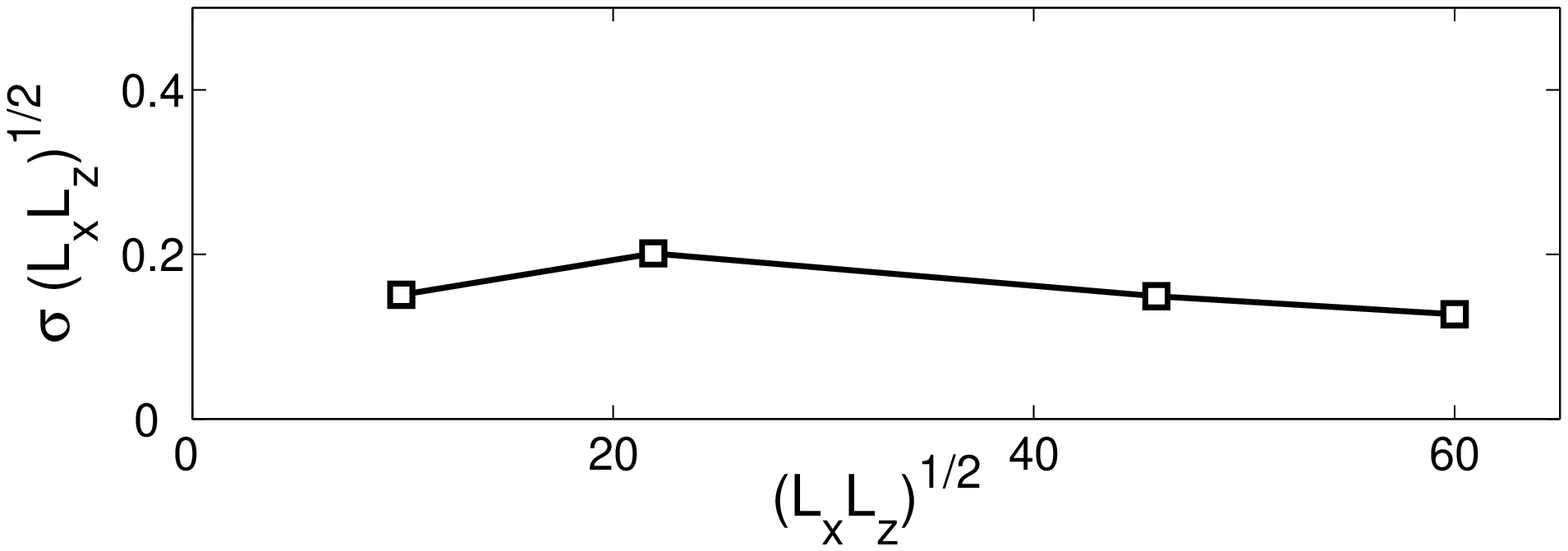}
\EC
\caption{Width $\sigma$ of the PDFs of $E$
compensated for the presumed extensive behavior of the featureless turbulent regime at $\RE=420$ for $L=16.9$, 35.3, 76.2, and 98.5, from data presented in Fig.~\ref{fig7}.\label{fig8}}
\EF
By eye, it can be seen in Fig.~\ref{fig7} that $\sigma$ decreases as $L$ increases, which can be understood as a consequence of averaging over wider domains.
Assuming local random behavior at the scale of the MFU, one can view the system as an assembly of independent sub-systems (which is of course not the case but may serve as a template), in which case one expects the standard deviation of the fluctuations to vary as the inverse square-root of their number, i.e. as the inverse of the square-root of its surface. Accordingly, in Fig.~\ref{fig8} we plot $\sigma$ compensated by its presumed variation with the size, i.e. $\sigma\sqrt{L_xL_z}$ as a function of $\sqrt{L_xL_z}$.
The fact this quantity is indeed approximately constant supports the underlying assumption but, by virtue of contrast, also suggests that it will have to be reexamined when the pattern will set in as a result of the interaction between MFUs at lower \RE.  

\subsection {Correlation lengths and the temporal--spatiotemporal issue\label{3.3}}

In order to gain further insight into the temporal$\,\to\,$spatiotemporal transition at a `microscopic' level, we now study the changes in the correlation functions upon domain size variation.
Here we focus on the streamwise perturbation velocity component obtained upon subtraction of the base flow profile, $\tilde u=u-y$. We consider the streamwise and spanwise time-averaged correlation functions  in the mid-plane $y=0$ where the base flow is strictly zero. They are defined as:
\BE
\label{eqn_CorCoeff}
C_x := \overline{\left( \frac{\langle \tilde u ({\bf x}+x\,\mathbf{\hat x},t)\tilde u({\bf x},t)
\rangle}{\langle \tilde u({\bf x},t)\tilde u({\bf x},t) \rangle}\right)}\,,\qquad
C_z :=  \overline{\left(\frac{\langle \tilde u ({\bf x}+z\,\mathbf{\hat z},t)\tilde u({\bf x},t)
\rangle}{\langle \tilde u({\bf x},t)\tilde u({\bf x},t) \rangle}\right)}\,,
\EE
where ${\bf x}$ is the in-plane running point while $\langle \;\rangle$ and the overlines denote averaging over space and time, respectively.
\BF
\BC
\includegraphics[width=0.46\textwidth]{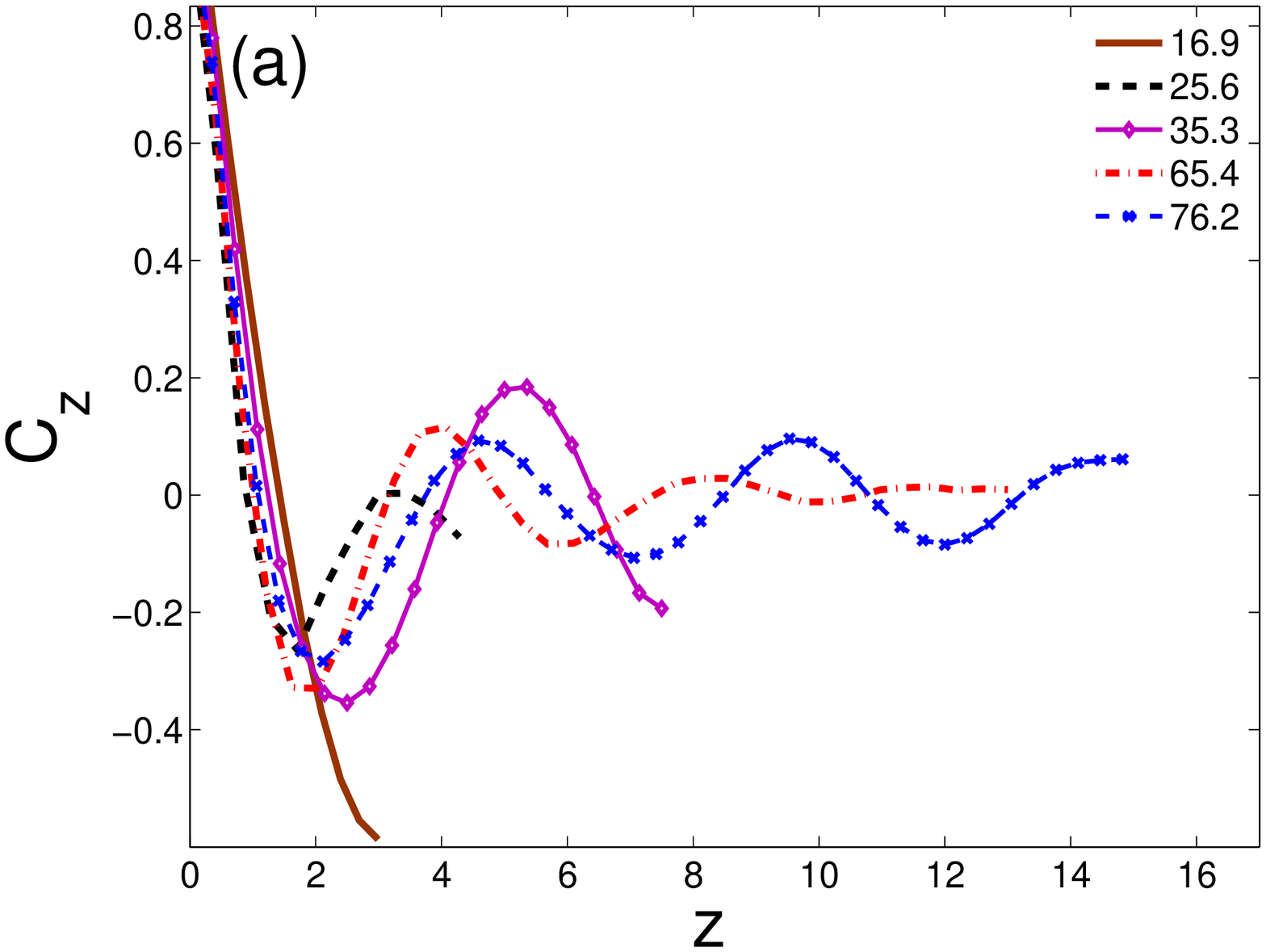}\hskip2em
\includegraphics[width=0.46\textwidth]{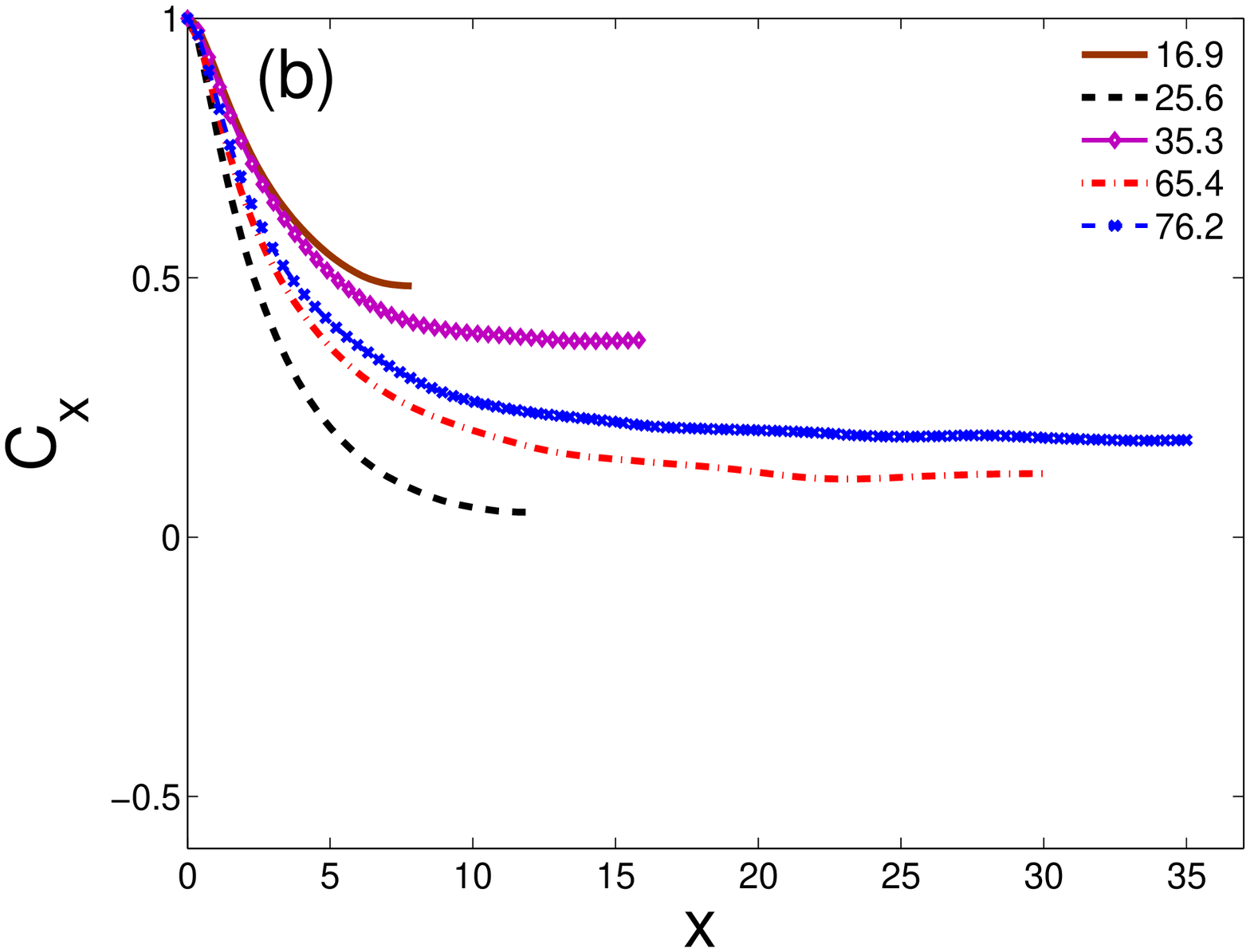}\\[2ex]
\includegraphics[width=0.46\textwidth]{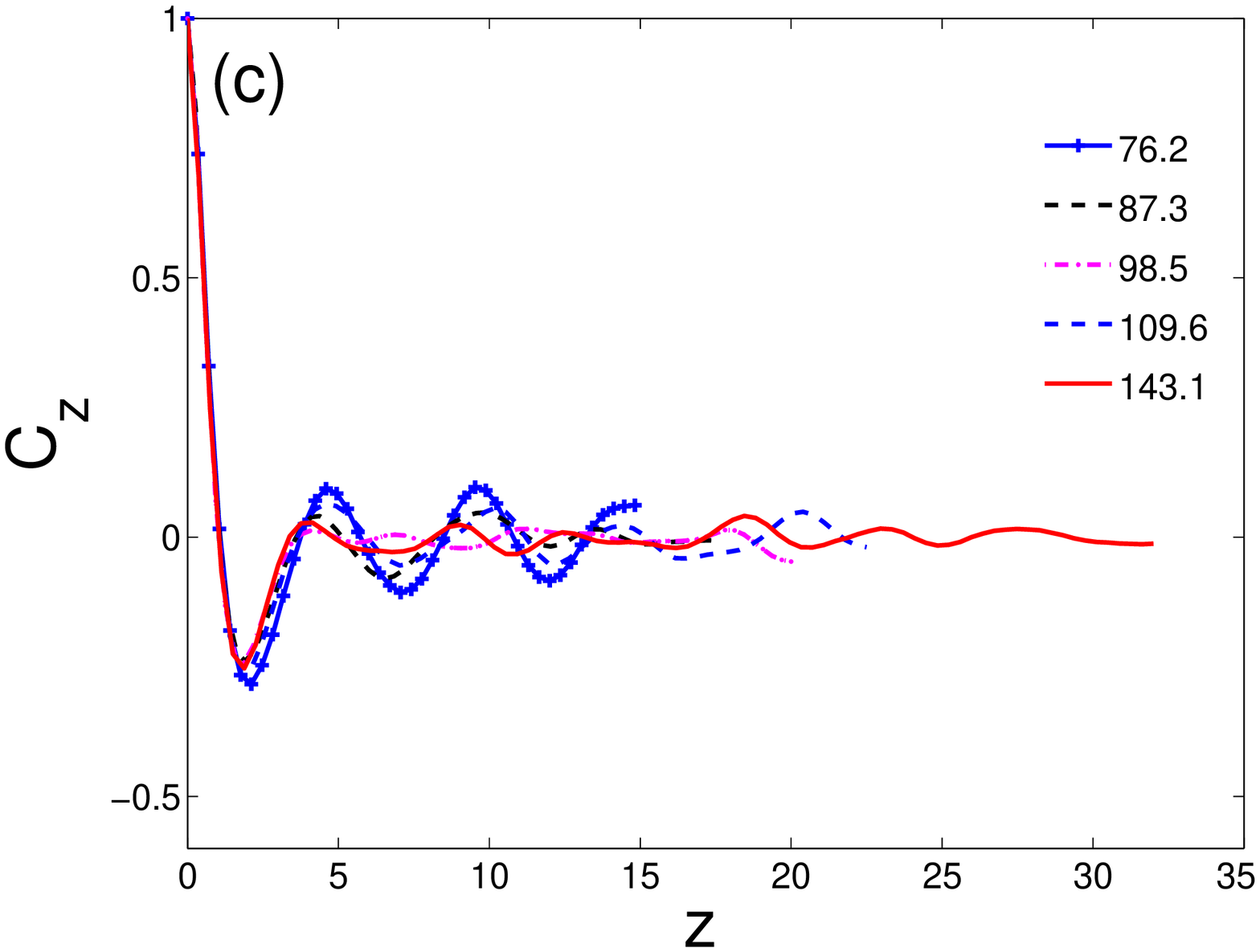}\hskip2em
\includegraphics[width=0.46\textwidth]{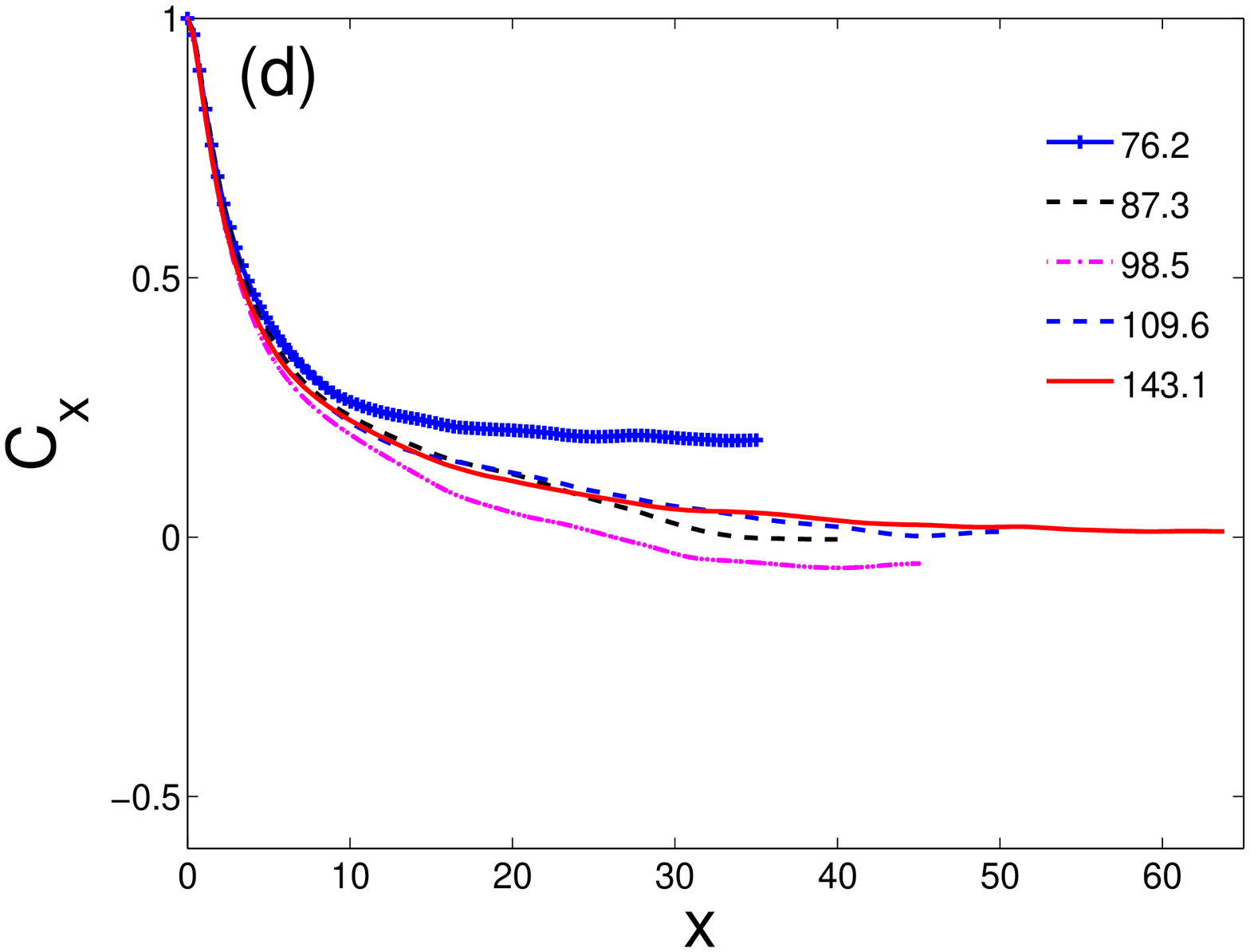}
\EC
\caption{(Color online) Time-averaged streamwise velocity correlations at $\RE=410$ for various domain sizes as indicated in the legends.
(a,c) Along the direction $\mathbf{\hat z}$.
(b,d) Along the streamwise direction $\mathbf{\hat x}$. Averaging over
1000 successive snapshots taken every $\delta t=1$
is performed.\label{fig9}}
\EF
Owing to the periodic boundary conditions in $z$ and $x$, the correlation functions are periodic with periods $L_x$ and $L_z$.
Furthermore, symmetry considerations lead to expect them to be symmetric with respect to the origin, which is approximately the case.
Accordingly, we only display one-sided symmetry-averaged correlations profiles over $[0,L_z/2]$ or $[0,L_x/2]$ for the different domains that we have considered.
They are presented for $\RE=410$ in Figure~\ref{fig9} which shows $C_z$ and $C_x$, in panels~(a,c) and~(b,d), respectively.

Due to the constraints brought by the boundary conditions, both $C_z$ and $C_x$ show that small domains are highly correlated.
It should however be noticed that $C_z$ passes zero even for the smallest domain, which is consistent with the presence of a pair of streaks (see Fig.~\ref{fig3}).
Information about processes at the scale of streak, such as the regeneration mechanism of turbulence can thus be studied in such domains.
That the process depends much more on the velocity profile variation in $z$ than in $x$ can be understood from the high level of correlation along $x$ (Fig.~\ref{fig9}b, $L$=16.9).
With increasing domain size $L$, $C_z$ oscillates with a period typical of the streaks. Oscillations seem to persist even at $L=143.1$. On the other hand, the streamwise correlation function $C_x$ decreases much more slowly and may reach zero, testifying for the very elongated character of these structures.

The above discussion shows that $L=76.2$--87.3 forms some kind of a boundary between the two distinct regions which exhibit temporal and spatiotemporal dynamics.
This boundary will be further explored in section \ref{3_3} below.
It is suggested that only domains able to accommodate these naturally existing elongated streaks are those that will exhibit banded patterns at lower \RE.
In this connection let us notice that, in their work, Barkley and Tuckerman were implicitly taking this feature into account by using their tilted domains since periodic conditions were correlating streaks that were shifted by one spanwise MFU width $\lambda_z$ so that their
streamwise amplitude could be modulated on a scale much longer than the period imposed by their pseudo-streamwise conditions at $L_{x'}=\lambda_z/\sin\theta$, where $\theta$ is the angle between their domain short direction and the $x$ axis \cite{BarTuc05}.
In the simulation corresponding to their Fig.~1, modulation took place over 9 periods, corresponding to simulations in a periodic box of size $L_x=97$ and $L_z=43$ aligned with the flow.
The limitation of their approach is that correlation in the spanwise direction is somewhat enhanced by the shift condition, which ends in patterns that look much more regular than what one obtains in boxes like our largest ones, $L=109$ ($L_x=100$, $L_z=45$, close
to their effective dimensions) or $L=143.1$ (see snapshots in Fig.~\ref{fig6}).

\subsection{Bifurcation diagram of transition in plane Couette flow\label{3_3}}

Bifurcation diagrams for PCF at different values of $L$ are displayed in Fig.~\ref{fig10}, left panel.
In the right panel we show them unfolded according to the system size, i.e. in a \RE--$L$--$D$ coordinate system, the presentation adopted for Fig.~\ref{fig2}.
\BF
\BC
\includegraphics[width=0.48\TW]{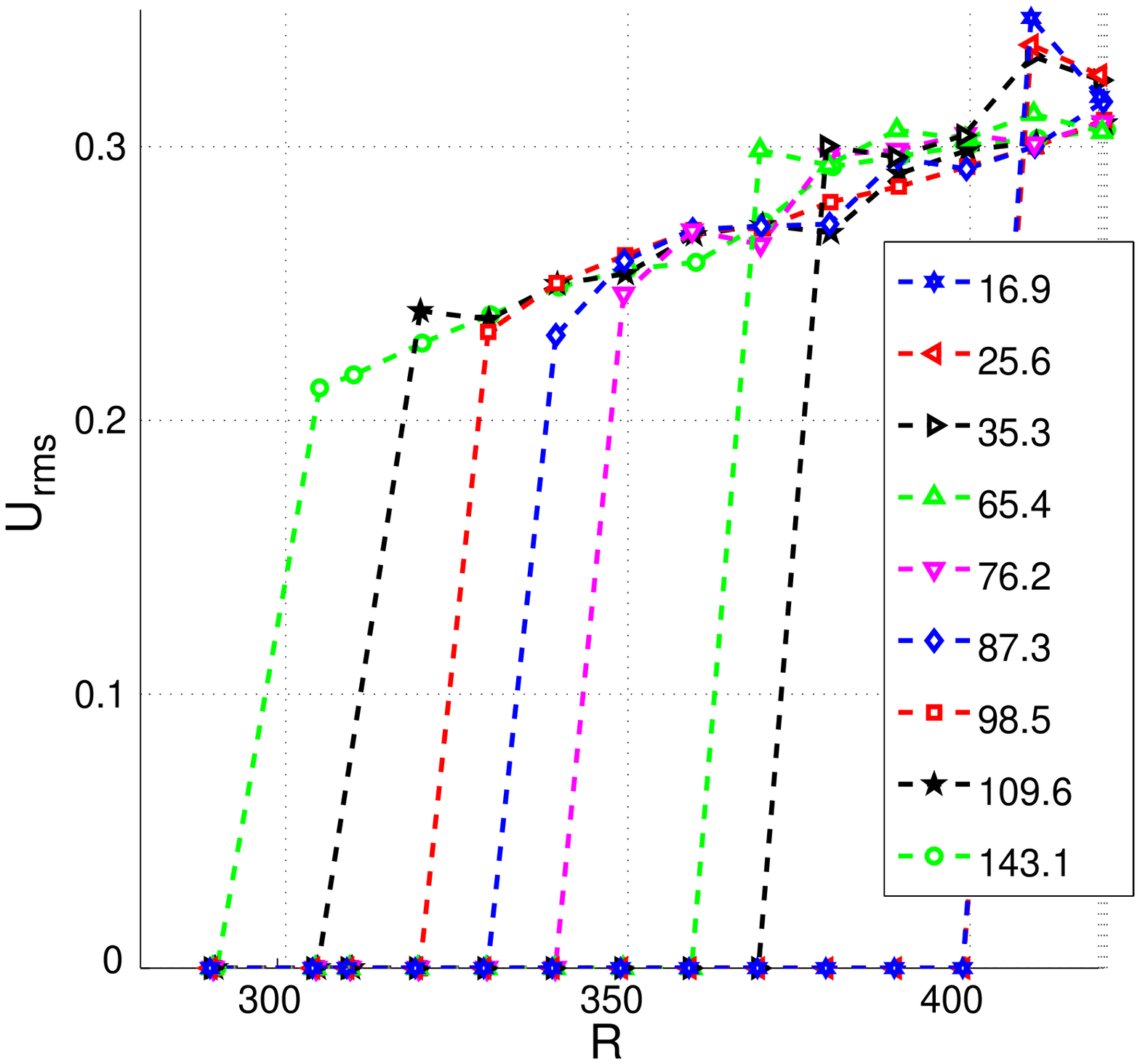}
\hfill
\includegraphics[width=0.48\TW]{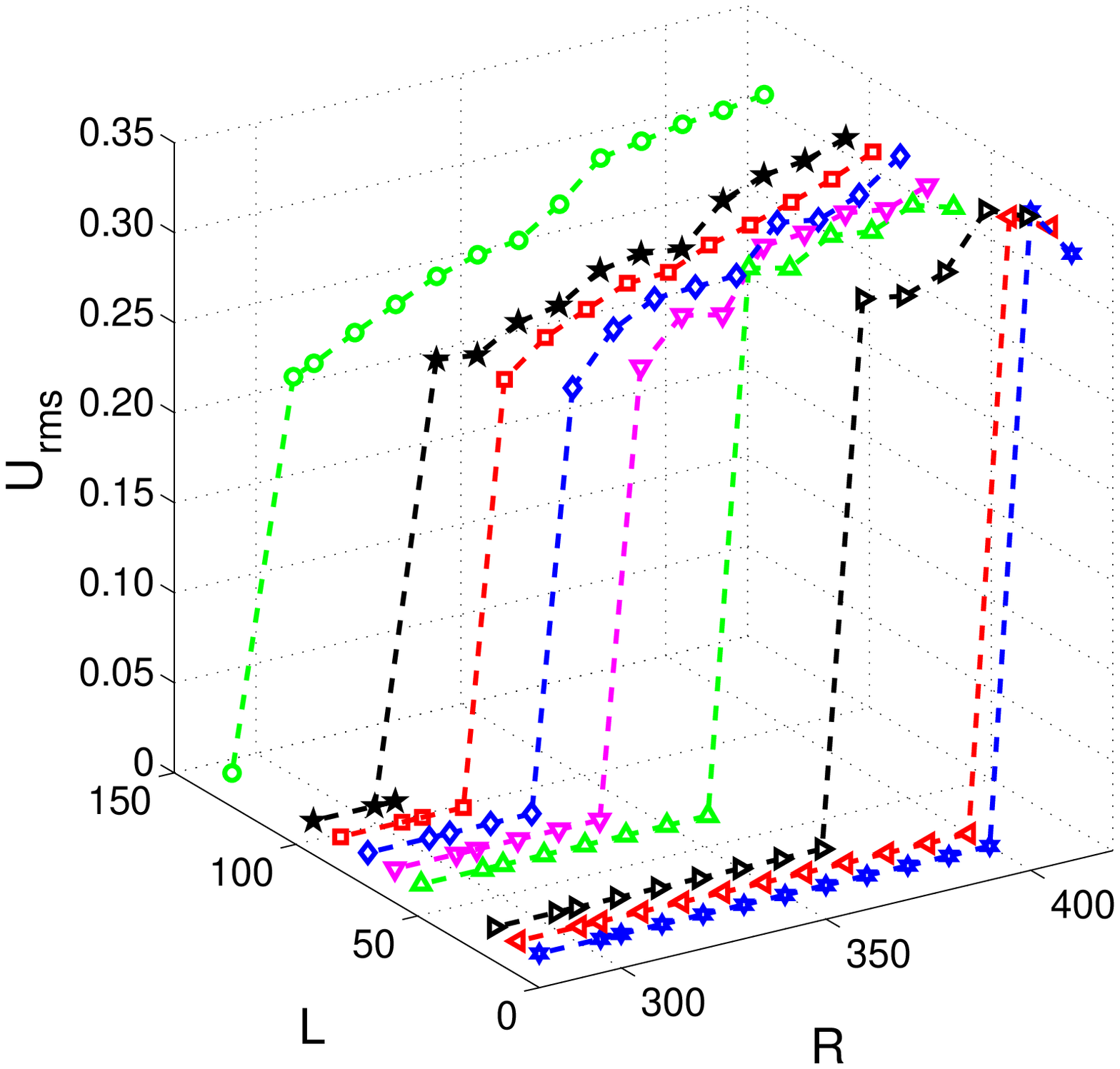}
\EC
\caption{(Color online) (a) Bifurcation diagrams in the \RE--$D$ plane for the different system sizes indicated in the legend, with $D\equiv\Urms$.
(b) Three-dimensional unfolding of (a) according to system size $L$.\label{fig10}}
\EF
The diagrams are all obtained following the same adiabatic protocol with $\Delta R=10$ and $\Delta T=10^3$, starting from states prepared at $\RE=420$, except for systems with $L=16.9$ and 25.6 that are started at $\RE=450$.
The distance $D$ used to characterize the state of the system is \Urms\ averaged over the last 600 time units.
In all cases, an abrupt transition from sustained turbulent regime to laminar flow is observed as \RE\ is decreased.
It should however be stressed that the exact quantitative position of the jump depends on the protocol:

$\bullet$\quad When $L$ is small, the transition is probabilistic \cite{EcFaScSc08,Sch_etal10} but a single trajectory is followed.
A better determination would necessitate to determine PDFs of transient lifetimes and the value of \RE\ at which the mean lifetime is larger than some value beyond which the flow regime would be considered as sustained.

$\bullet$\quad At larger $L$, the relaxation of turbulence has also probabilistic features and choosing $\Delta T$ too small may not ensure us that the system has explored a sufficiently large part of its accessible phase space so that the result is still sensitive to the initial state at the value of \RE\ considered, which is the state at the end of the simulation at $\RE+\Delta\RE$.
On the other hand, when $\Delta\RE$ is too large, setting the system at \RE\ from $\RE+\Delta\RE$ can be a large perturbation which may place the initial condition outside the attraction basin of the turbulent state at the considered \RE, leading to an overestimation of the \RE\ corresponding to the jump in the bifurcation diagram for that $L$.
This case corresponds to the quench experiments in which the initial state is systematically taken to be fully turbulent \cite{Betal98}.
The risk is more limited if $\Delta\RE$ is small enough as in the adiabatic protocol that we are following.

We are confident that the decrease of the Reynolds number at which decay to laminar flow occurs seen in Fig.~\ref{fig10} as $L$ increases is not an artifact of the protocol.
This decrease in \RE\ means that, when $L$ is increased, the flow needs to be forced less vigorously to remain turbulent because its effective phase space has enlarged so that it has more freedom to evolve non-trivially in a spatiotemporal manner, rather than strictly temporally when it is confined by the lateral boundary conditions at small distances.
The increase in the region of non-laminar flow with increasing domain size can also be inferred from the plot in Fig.~\ref{fig11} where colors from blue$\,=\,$zero to red shows \Urms\ by increasing values.
Also shown in this figure
\BF
\BC
\includegraphics[width=\TW]{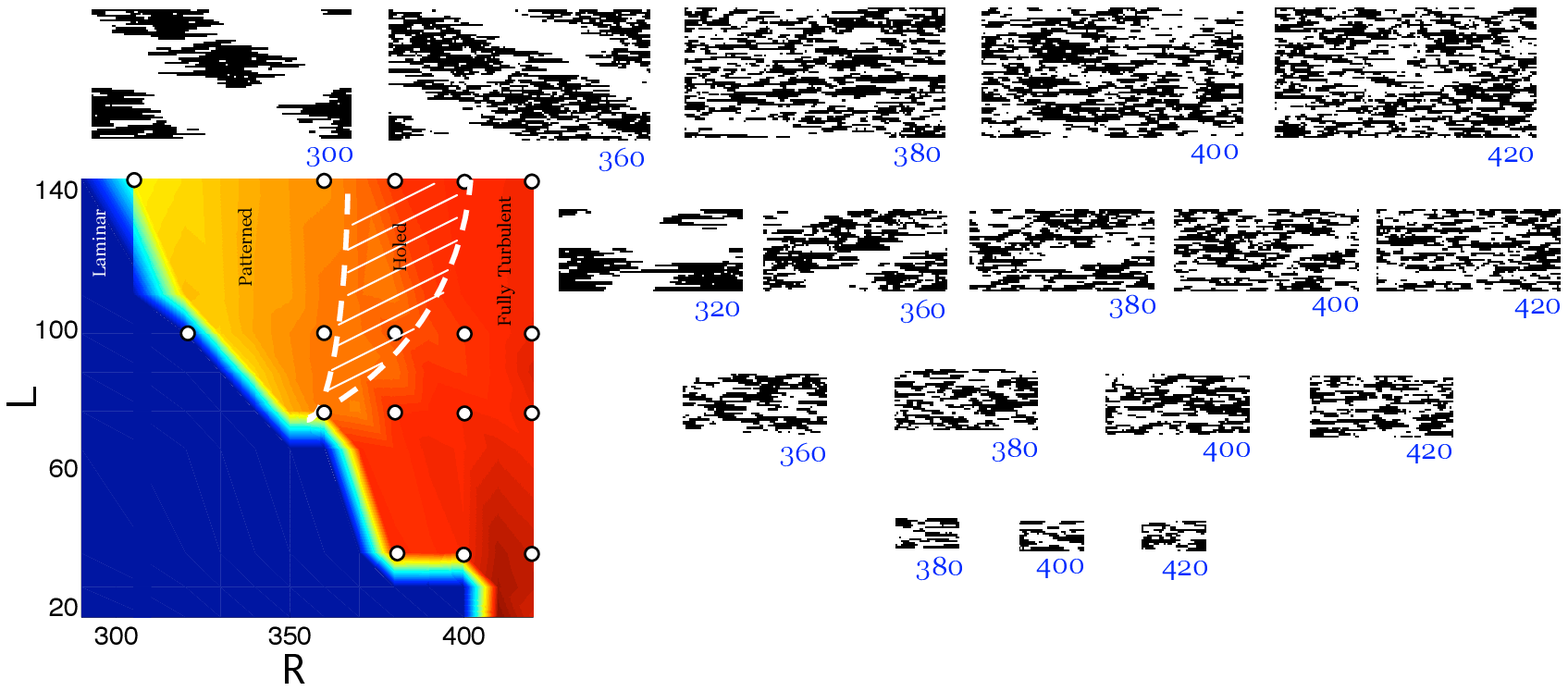}
\EC
\caption{Left: (Color online) Bifurcation diagram in the \RE--$L$ plane with color coded
values of $D\equiv\Urms$; red (rendered medium gray) correspond to the largest value of $D$ (turbulent at large $L$, chaotic at small $L$), blue (rendered dark gray) is for laminar flow where $D=0$, patterns and rigged states have intermediate values of $D$ from yellow to orange (rendered light to medium gray).
Above and on the right: BW representation of turbulent-laminar patches for $L=143.1$ (top), 98.5, 76.2 and 35.3 (bottom).
Corresponding Reynolds numbers are indicated at the bottom right corner of each image. White dots in the left panel show the location of the states corresponding to the BW plots.
Cases ($L=143.1$, $\RE=300$) and ($L=98.5$, $\RE=320$) are decaying and eventually become laminar.\label{fig11}}
\EF
are black-and-white (BW) coded snapshots of the system for a series of sizes and Reynolds numbers.
The corresponding cases are indicated by white dots in the left panel.
These images are obtained by coarse-graining $\|\mathbf{\tilde v}\|^2$ from the final state of each simulation at the corresponding \RE\  in cells of size $1\times1\times1$, following a procedure described in \cite{RoMa10}.
The top ($y\geq0$) and bottom ($y\leq0$) cells are separately coarse-grained to take the lateral shifts of the laminar--turbulent regions in both halves into account \cite{CoVa66,BarTuc07,RoMa10}.
BW thresholding is then performed, a pixel being termed `turbulent' and B-colored when the mean of the energy of the two corresponding stacked cells is larger than half the average energy in the whole system, `laminar' and W-colored otherwise.
This automated cut-off criterion has been found suitable all across the ranges of \RE\ and domain sizes studied, yielding pictures visually similar to their $y$-averaged, non-coarse-grained counterparts.

The observation of the bifurcation diagrams in Fig.~\ref{fig10} (left) does not lend itself to clearly differentiate the regions where well-oriented patterns occur and where they do not, though for domains $L=65.4$ and smaller, \Urms\ is seen not to vary much before decay, whereas in larger domains a gradual change is observed.
However, the BW energy plots in Fig.~\ref{fig11}, e.g. $(L=143.1,\RE=400$) or ($L=98.5,\RE=380$), show that apart from the patterned region, there exists a parameter region where the surface of the system is `moth-eaten' or `riddled' with fluctuating laminar troughs.
This can also be realized from the changing magnitude of \Urms\ in the color plot if its slow decrease with \RE\ is attributed to a concomitant increase of the laminar fraction.
A hatched domain limited by a white dashed line in the left part of Fig.~\ref{fig11} roughly indicates the presence of these riddled states. This particular regime thus appears as a precursor to the banded regime when the patterns are approached by increasing $L$ or decreasing \RE, situated in the bifurcation diagram like the intermittent regime described by Prigent {\it et al.} \cite{prigent03} and Barkley \& Tuckerman \cite{BarTuc05} and to which it corresponds closely. A possible equivalent  of it in pipe flow seems to be the intermittent laminar-flash regime described in \cite{MoxBar10} near the transition to uniform turbulence that we call featureless.
In our numerical configuration, the bifurcation diagram reaches its eventual large aspect-ratio limit `featureless--riddled--banded--laminar'  for systems of sizes beyond $L=76.2$.

Further visualization in Fig.~\ref{fig12}
\BF
\BC
\includegraphics[width=0.2461\TW,clip]{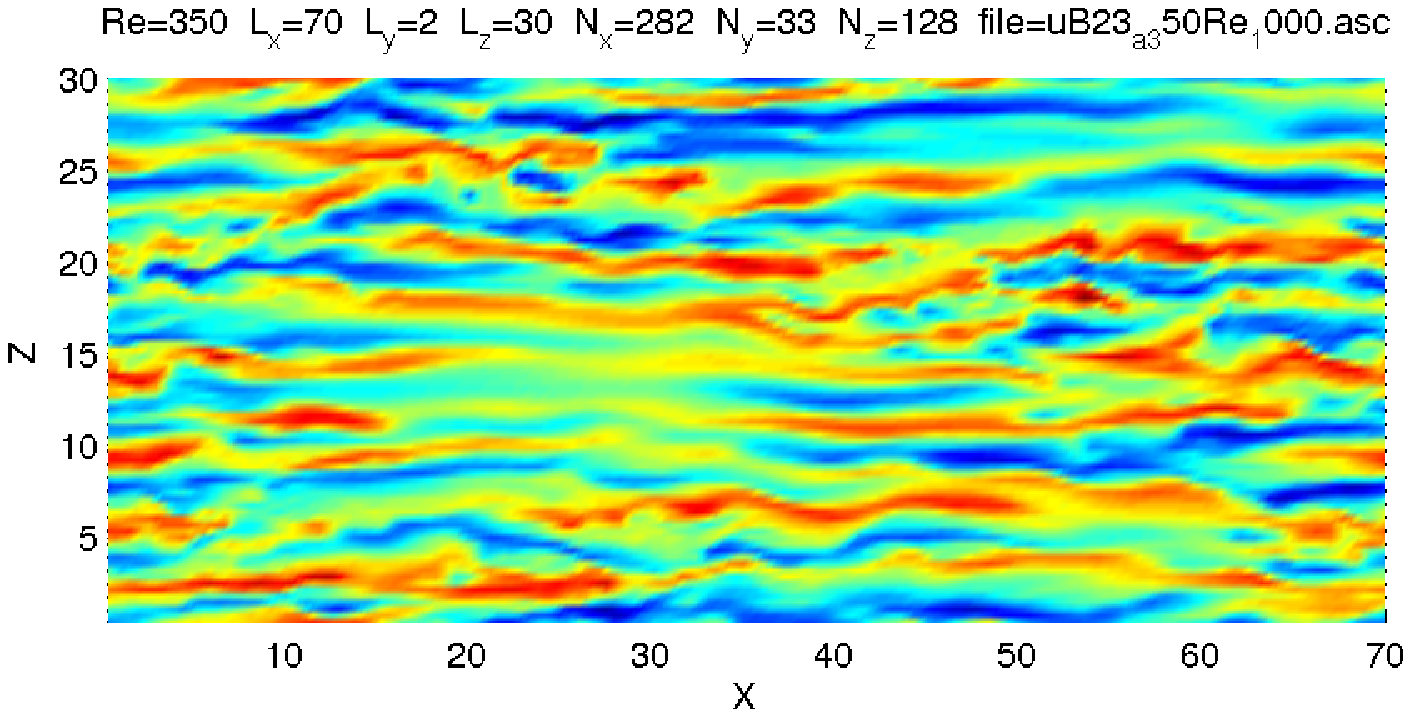}\\[4ex]
\includegraphics[width=0.28125\TW,clip]{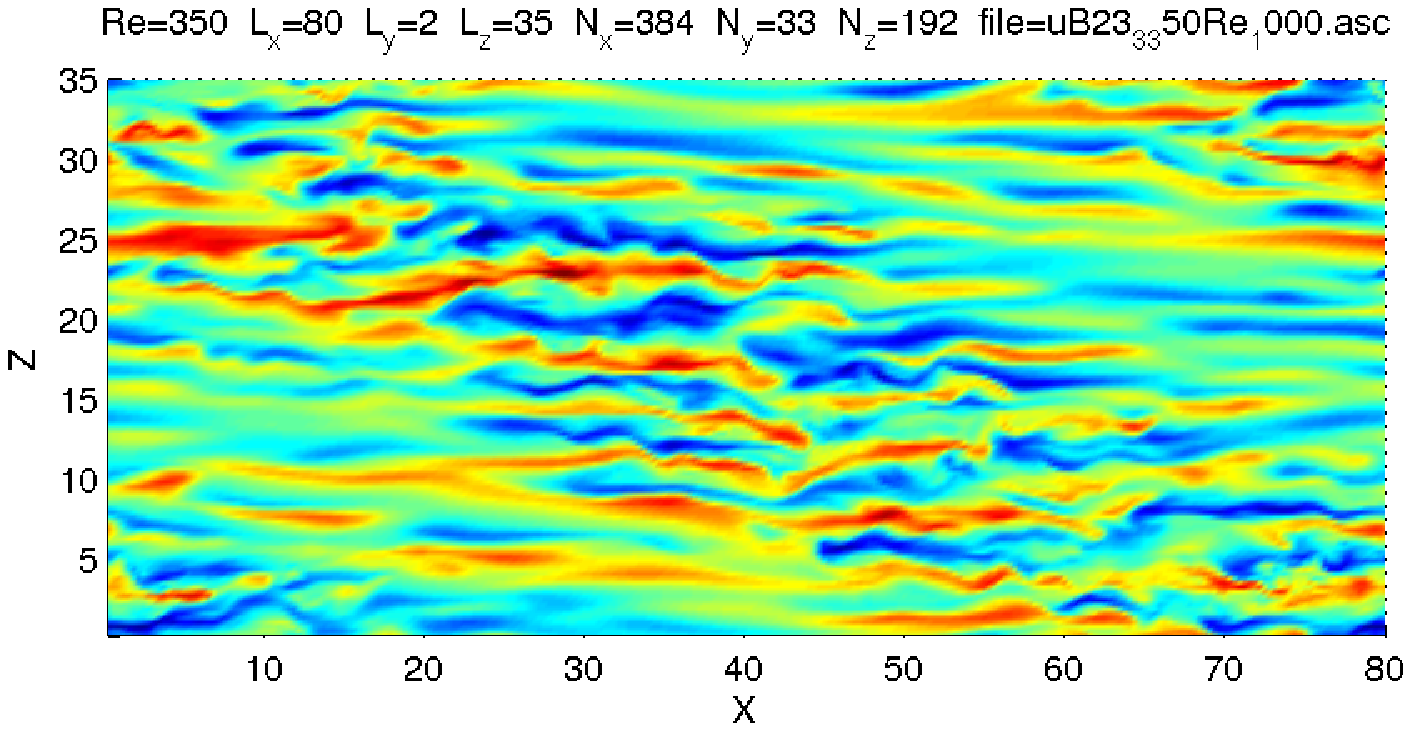}
\hspace{0.05\TW}
\includegraphics[width=0.3164\TW,clip]{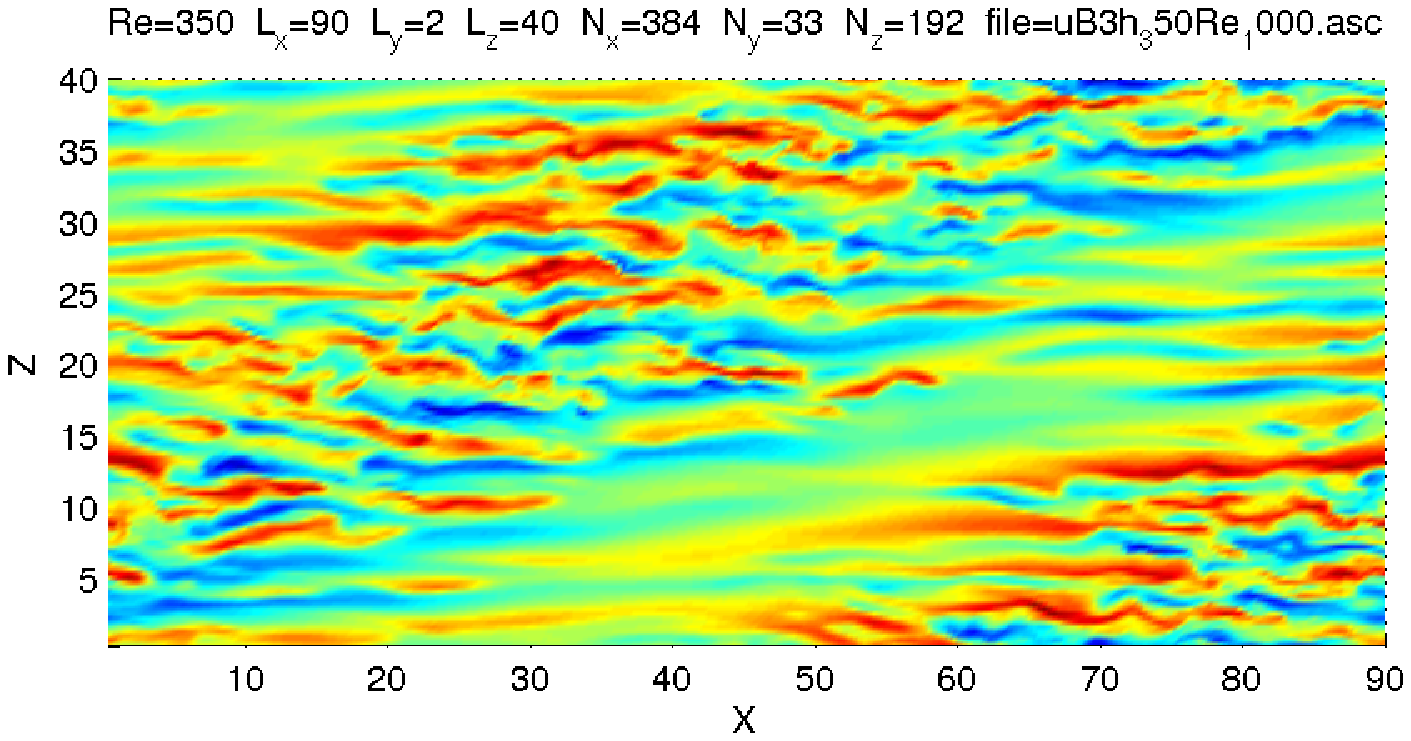}\\[4ex]
\includegraphics[width=0.3515625\TW,clip]{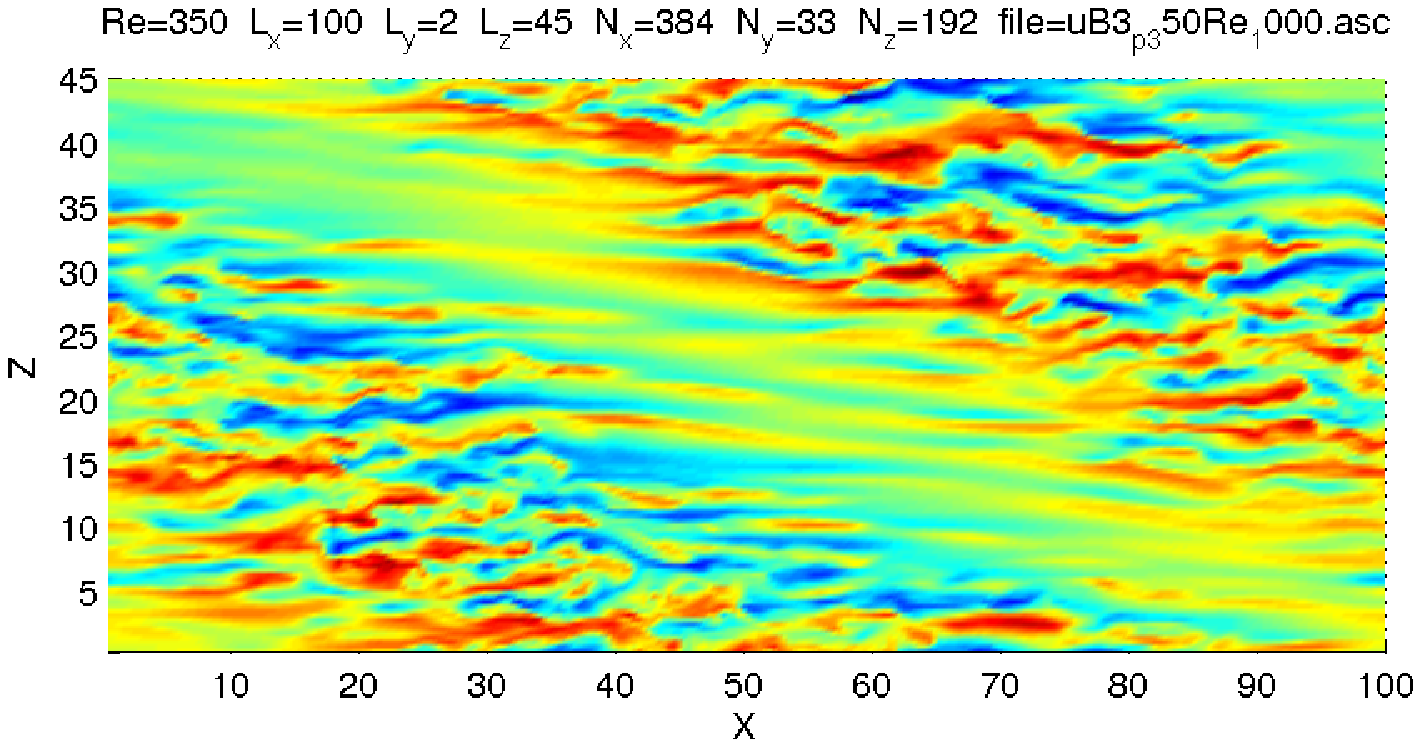}
\hspace{0.05\TW}
\includegraphics[width=0.45\TW,clip]{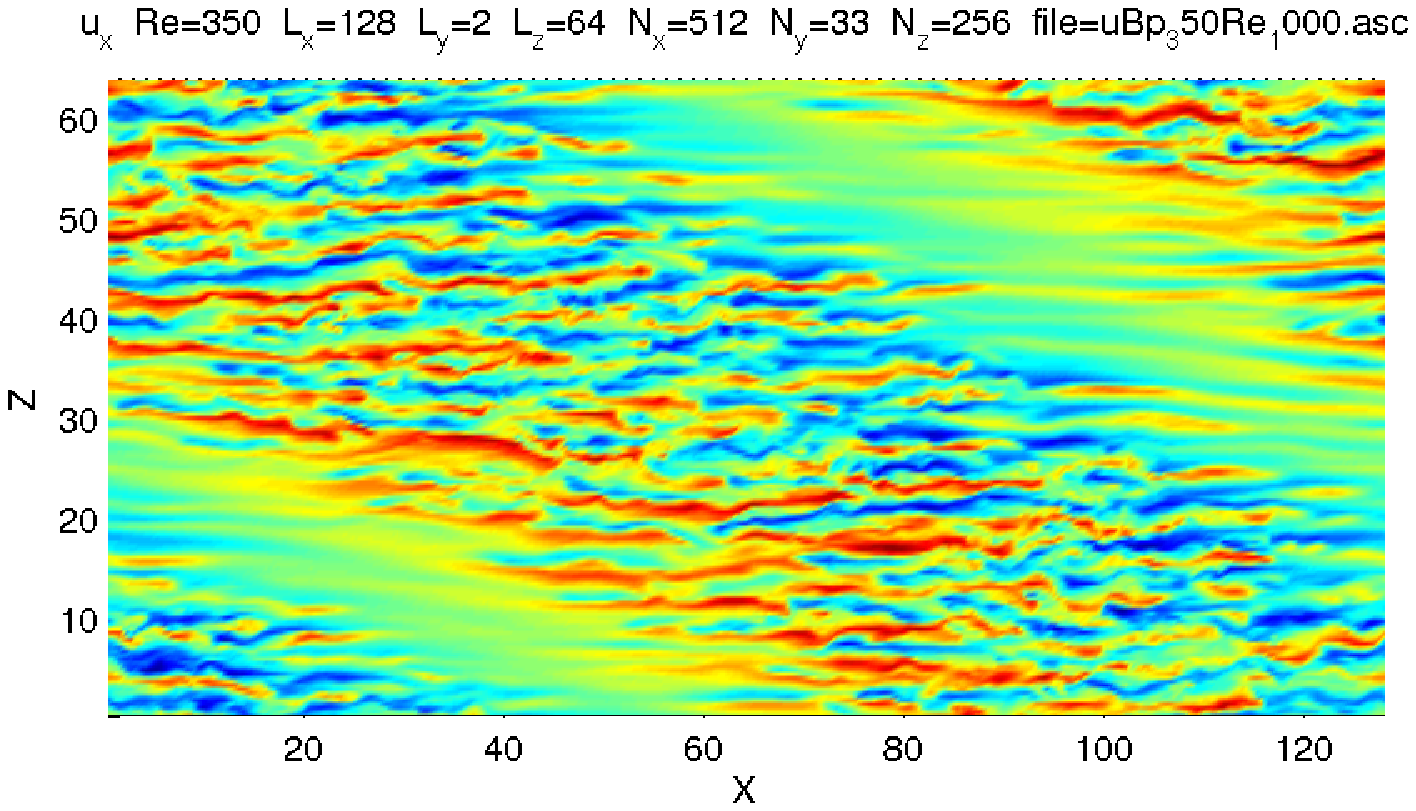}
\EC
\caption{(Color online) Streamwise velocity fluctuation $u$ in the plane $y=0$
for different domains at $\RE=350$: $L=76.2$, ($\theta=23.2^\circ$, top), 87.3 ($\theta=23.6^\circ$, center-left), 98.5 ($\theta=24.0^\circ$, center-right), 109.6 ($\theta=24.2^\circ$, bottom-left), 143.1 ($\theta=26.6^\circ$, bottom-right).
\label{fig12} }
\EF
shows systems with $L\ge76.2$ where one can visually observe the tendency to form a well-organised pattern as $L$ is increased.
Below this domain size, i.e. $L\le65.4$, no pattern or troughs could be observed upon reducing \RE\ down to the final laminar state.
Accordingly, patterns seems to need systems larger than some minimal size to exist.
However, this size should rather be considered to mark some crossover than to be `critical' in the usual sense because, there cannot be any decisive criterion to decide whether one has
more or less oblique, elongated, laminar troughs like for $L=76.2$, or already well-formed bands, like for $L=87.3$.
As noted above, these laminar troughs are also found in very large domains at higher \RE\ and are precursors to the bands when, at given $L$, \RE\ is reduced adiabatically from well above \RT.
In fact, from the plots in Fig.~\ref{fig10} (left), one can see that  $\RE=350$ is close to the transition to laminar flow for $L=76.2$ and 87.3, hence at the corner of the white hatched region in Fig.~\ref{fig11}.
Systems in the crossover region thus go from turbulent to laminar by skipping the patterned stage but still going through the intermittent-riddled regime.

\section {Summary and Conclusions\label{4}}

The formation of banded patterns in flows such as plane Couette flow, circular Couette flow and plane Poiseuille flow is typically a spatiotemporal problem.
And true enough, this phenomenon has been experimentally observed in domains with lateral dimensions that are more than two orders of magnitude larger than the wall-normal distance.
Numerical simulations of the flow field in these large systems have become possible only recently, thanks to the less stringent resolution requirements to resolve all the scales of low-\RE\ turbulent flow and to our ever increasing computational capabilities.
Despite these favorable circumstances, simulations cannot be performed over long enough durations to bring definite answers about the statistics of the transitional regime in the large aspect ratio limit.
Previously, with fully developed wall-bounded turbulence in mind (high-\RE), numerical
computations and theory were more focused on small domains.
In the moderate-\RE\ range of interest to the problem of transition, these small systems are best analyzed within the framework of finite-dimensional dynamical systems theory.
The present work stays in between these two types of studies, wherein we start with low-\RE\ turbulence in small domains where temporal dynamics takes place, and increase the domain size gradually to reach the spatiotemporal regime.
An adiabatic decrease of \RE\ for each domain then gives information about the different regimes visited by the system from featureless turbulence to laminar flow.

Upon decreasing \RE, starting from a featureless fully turbulent flow, we have obtained that, in small systems, the mean value of the perturbation energy remains roughly constant as a function of \RE\ down to the point where the chaotic regime breaks down into laminar flow, whereas it is regularly decreasing in larger systems.
Furthermore, visualization of the flow field indicates that the mean-energy variation corresponds to the system entering a spatiotemporal regime that at first (high \RE) presents fluctuating laminar troughs and later (low \RE) steady oblique patterns of bands alternately laminar and turbulent, but absent in the small systems in which spatial coherence implies a mostly temporal dynamics.

In the featureless regime, fluctuations of the mean perturbation energy have distribution probabilities with a single marked hump around the mean for all domain sizes.
Their variance is decreasing with increasing system size, and fluctuations in the widest domains are close to normal, which can be understood as a result of the additivity of local
(MFU scale) fluctuations that remain little-correlated as long as the system is sufficiently far from the laminar breakdown, expressing the extensive character of the featureless regime.

A study of the spanwise  and streamwise dependence of the streamwise velocity component correlation function as functions of the system size has also been performed.
The spanwise variation accounts for the self-sustaining process by displaying a periodic
dependence at the scale expected for the streaks.
More importantly, the slow streamwise variation expresses correlations reinforced by the periodic boundary conditions in the smallest systems.
The ability to take into account this streamwise dependance properly seems an important ingredient in order to explain the absence of patterning noticed in these systems.
Indirectly, this also explain the observation of the pattern in the Barkley--Tuckerman tilted but short domains since their domain width was chosen to fulfill a commensurability condition ensuring the periodic continuation of streaks at the tilt angle observed in the experiments, thus mimicking longer domains in the streamwise direction.
In our computations, patterns were found for $L>76.2$, i.e. $L_x>70$.

Inspection of flow fields (e.g. Fig.~\ref{fig12} or the BW insets in Figure~\ref{fig11}) indicates that, in accordance with the information gained from correlation functions, one can only observe intermittent laminar troughs for $L=76.2$, and see any organized pattern of laminar--turbulent oblique bands only for $L=87.3$ and beyond.

In Figure~\ref{fig2}, a conjecture was presented about the bifurcation diagram of PCF, connecting systems of MFU size exhibiting temporal dynamics \cite{Sch_etal10} to large aspect-ratios systems displaying spatiotemporal dynamics and patterns \cite{prigent03}.
Substantiating this conjecture at a quantitative level, Figures~\ref{fig10} and \ref{fig11} display the bifurcation diagrams obtained for various system sizes.
The smallest domains follow the direct route `[chaotic flow]$\>\to\>$[laminar flow]' ({\it via\/} chaotic transients), whereas in very large domains one has `[turbulent flow]$\>\to\>$[riddled regime]$\>\to\>$[oblique pattern]$\>\to\>$ [laminar flow]'.
For $L$ in the range 75--85 the [oblique pattern] stage is skipped, which marks the crossover from temporal to spatiotemporal dynamics.
Furthermore, the threshold at which the pattern decays, \RG, is shown to decreases with increasing system size, seemingly tending to a constant in the large aspect-ratio limit.
A precise quantitative estimate of this limit was however outside the scope of this paper since the final turbulence breakdown still keeps probabilistic features that requires statistics (longer time series, large number of independent realizations), especially in what regards the occurrence of turbulent patches issued fragmented bands turning into turbulent spots close to \RG.
At any rate the exact value is of little interest, all the more that periodic boundary conditions tend to stabilize the pattern, which thus artificially decreases \RG\ at moderate aspect ratios.
The observed trends (decrease of \RG\ as $L$ is increases, shrinking of the band regime to the benefit of the intermittent riddled regime) however go in the same direction as the experimental findings showing that at intermediate aspect ratio, the transitional range is pushed at higher \RE\ and that patterns can hardly be observed---see \cite{TiAl92} and Bottin's thesis
\cite[\S4.4]{Bo98}---though no quantitative link can be made in view of the differences in the lateral boundary conditions.

All in all, the recognition of a crossover size beyond which the transition to/from turbulence in PCF is undoubtedly a spatiotemporal process, and the need of a faithful account of streamwise correlations at this scale, either directly or indirectly {\it via\/} the tilted-domain trick, seem our most important observations.
They might help us to unravel the physical mechanism behind the organization of low-\RE\ turbulence in wall-bounded flows, which still largely remains an enigma.

\bibliographystyle{plain}
\bibliography{pre_arxiv}

\end{document}